\documentclass[prl,twocolumn,amsmath,amssymb,superscriptaddress]{revtex4-1}
\usepackage{epsfig, graphicx,graphics,amsmath,amssymb,float}
\usepackage[T1]{fontenc}
\usepackage[latin9]{inputenc}
\usepackage{amsmath}
\usepackage{amssymb}

\usepackage{amscd}
\usepackage{bm}
\usepackage{psfrag}

\usepackage{bbm} 

\usepackage[caption=false]{subfig}
\usepackage{subfig}
\usepackage{color}
\usepackage[bookmarks=true,colorlinks,linkcolor=blue,urlcolor=blue,citecolor=blue]{hyperref}

\newcommand{\be}{\begin{equation}}
\newcommand{\ee}{\end{equation}}
\newcommand{\bea}{\begin{eqnarray}}
\newcommand{\eea}{\end{eqnarray}}

\newcommand{\mc}{\mathcal}
\newcommand{\mb}{\mathbf}

\begin{document}

\title{Spin chain network construction of chiral spin liquids}
\author{Gabriel  Ferraz}
\affiliation{International Institute of Physics,  Universidade Federal do Rio Grande do Norte, 
Natal, RN, 59078-970, Brazil}
\affiliation{Departamento de F\'isica Te\'orica
e Experimental, Universidade Federal do Rio Grande do Norte, 
Natal, RN, 59078-970, Brazil}
\author{Fl\'{a}via B. Ramos}
\affiliation{International Institute of Physics,  Universidade Federal do Rio Grande do Norte, 
Natal, RN, 59078-970, Brazil}
\author{Reinhold Egger}
\affiliation{Institut f\"ur Theoretische Physik,
Heinrich-Heine-Universit\"at, D-40225  D\"usseldorf, Germany}
\author{Rodrigo G. Pereira}
\affiliation{International Institute of Physics,  Universidade Federal do Rio Grande do Norte, 
Natal, RN, 59078-970, Brazil}
\affiliation{Departamento de F\'isica Te\'orica
e Experimental, Universidade Federal do Rio Grande do Norte, 
Natal, RN, 59078-970, Brazil}


\begin{abstract}
We show that a  honeycomb lattice of Heisenberg spin-$1/2$ chains with three-spin junction interactions allows for controlled 
analytical studies of chiral spin liquids  (CSLs).  Tuning these interactions to a chiral fixed point, we find a Kalmeyer-Laughlin CSL phase which here is connected to the critical point of a boundary conformal field theory.  Our construction directly yields a quantized spin Hall conductance and localized spinons with semionic statistics as elementary excitations.  We also outline the phase diagram away from the chiral point where spinons may condense. 
Generalizations of our approach can provide microscopic realizations for many other CSLs.
\end{abstract}
\maketitle

\emph{Introduction.---}Chiral spin liquids occupy a prominent position among the most exotic quantum phases of matter \cite{Wen2017}. 
As examples of quantum spin liquids \cite{Savary2016,Knolle2019}, they occur in magnetic insulators 
with  long-range-entangled ground states that break time-reversal and reflection symmetries. The historically first proposal is the Kalmeyer-Laughlin CSL  \cite{Kalmeyer1987,Wen1989}, a topological phase of interacting spins equivalent to a bosonic fractional quantum Hall state. 
The non-Abelian phase of Kitaev's honeycomb model in a magnetic field provides another CSL example, with Ising anyons as elementary excitations \cite{Kitaev2006}. 
Recent experiments have reported a quantized thermal Hall conductance for the Kitaev material $\alpha$-RuCl$_3$ \cite{Kasahara2018}, 
compatible with the chiral Majorana edge mode expected for this CSL phase.  
Various other CSL phases have been theoretically investigated \cite{Greiter2009,Chua2011,YaoLee2011,Bauer2014,He2014,Gorohovsky2015,Meng2015,Huang2016,Lecheminant2017,Kumar2015,Sedrakyan2015,Poilblanc2015,Hickey2016,Wietek2017,Yao2018} and are actively searched for in experiments, including gapless CSLs with spinon Fermi surfaces \cite{Fak2012,Bieri2015,Pereira2018,Bauer2019}. 

A major obstacle to the theory of CSLs comes from the shortage of analytical methods able to predict their occurrence and their physical properties in  
microscopic models. Apart from exactly solvable models \cite{Kitaev2006,Chua2011,YaoLee2011}, 
standard approaches employ parton mean-field theories that fractionalize the spin operator into fermionic or bosonic quasiparticles \cite{Wen1989,FradkinBook}, or use variational wave functions obtained by a Gutzwiller projection scheme \cite{Savary2016}. Such approaches are often able to capture the basic phenomenology when the CSL phase is indeed realized. However, since they rely on uncontrolled approximations, their predictions are often questionable, e.g., due to the neglect of interactions mediated by emergent gauge fields.  
In this Letter we establish a connection between chiral fixed points of boundary conformal field theory (BCFT) \cite{Cardy1986,Oshikawa2006} and CSL phases, and use it to formulate 
a controlled analytical construction scheme for CSLs where chiral junctions of multiple spin chains serve as the elementary building blocks in two-dimensional (2D) networks of spin chains.  Our approach markedly differs from standard coupled-wire constructions \cite{Kane2002,Fuji2019,Gorohovsky2015,Meng2015,Huang2016,Lecheminant2017,Tikhonov2019}, where CSL phases are studied for parallel chain models. In such schemes, one usually subjects the bosonized theory to a  
renormalization group (RG) analysis, where selected couplings flow to strong coupling.  For the dominant coupling, one then pins the corresponding 
boson fields by means of a semi-classical analysis of the respective cosine terms.
However, this procedure works at best for gapped CSLs only, and due to the presence of competing instabilities, it is 
difficult to reliably predict the location of the CSL phase in the parameter space of the microscopic model. 
Finally, in contrast to the spatial anisotropy inherent to parallel wire models, our chiral-junction network construction naturally preserves point group symmetries. 
This point may play an important role in protecting gapless CSLs \cite{Pereira2018,Bauer2019} and interacting topological crystalline  phases \cite{Song2017,Huang2017}.  

\begin{figure}[t]
\begin{center}
\subfloat[\label{fig1a}]{\includegraphics[width=0.35\columnwidth]{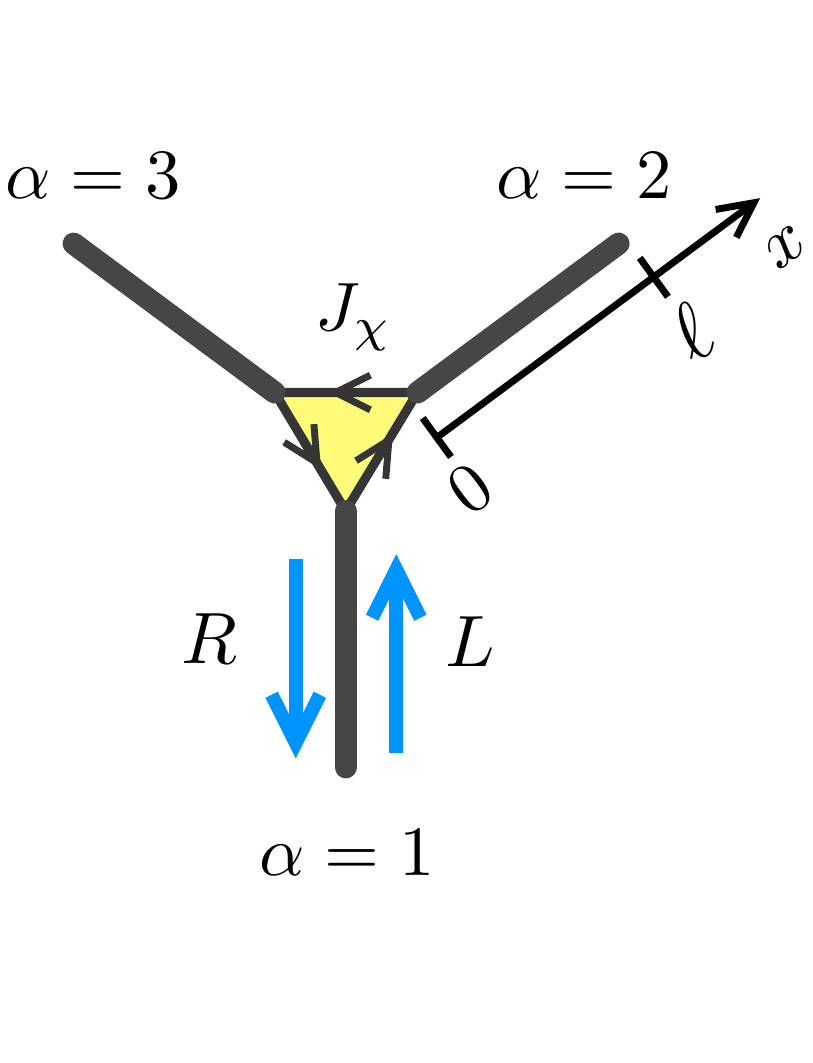}}\qquad
\subfloat[\label{fig1b}]{\includegraphics[width=0.55\columnwidth]{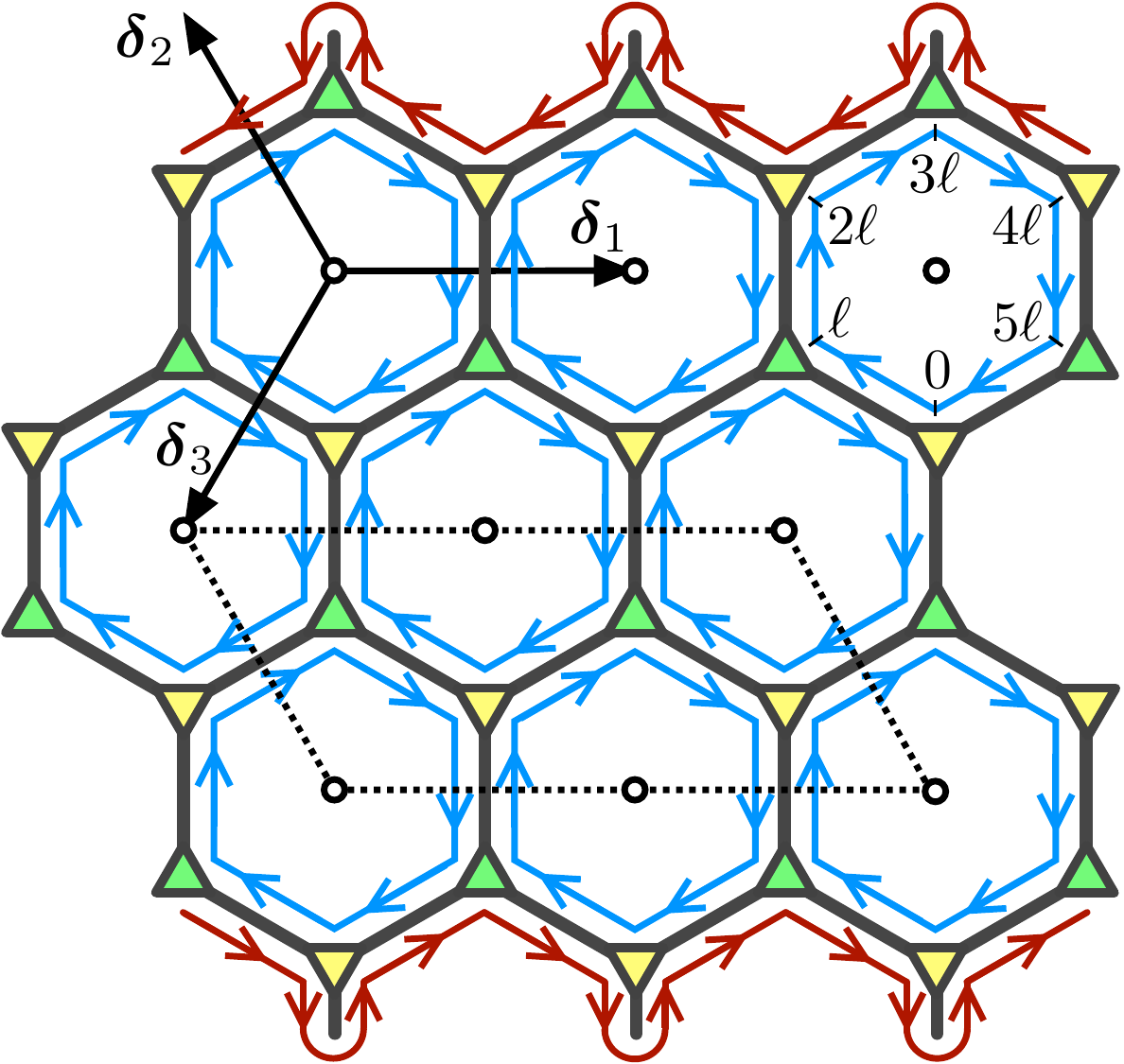}
}
\caption{\label{fig1} Y junction and 2D honeycomb network of spin-1/2 chains. (a) Single Y junction: Three Heisenberg chains of equal length $\ell\gg 1$ (with lattice spacing $a=1$) are
labeled by $\alpha=1,2,3$. They are coupled at the junction ($x=0$) by the three-spin boundary coupling $J_\chi$ in Eq.~\eqref{model}. At the chiral fixed point, $J_\chi=J_\chi^c$, incoming
(left-moving, $L$) spin currents are perfectly rerouted in a counterclockwise (for $J_\chi>0$) sense \cite{Buccheri2018,Buccheri2019}.
(b)  Adding Y junctions at each chain boundary $x=\ell$, and iterating the process, one obtains the shown 2D spin chain network.  Down-pointing triangles (yellow) mark the positions $\mb V$.
The neighboring junctions correspond to up-pointing triangles (green). The three vectors $\boldsymbol\delta_\alpha$ in Eq.~\eqref{bc2} are also indicated \cite{foot1}.  For $J_\chi=J_\chi^c$, 
the field theory reduces to decoupled chiral bosons describing circulating loop spin currents (blue), parametrized by the hexagon centers $\mb R$ (open circles) and the 1D coordinate $\xi\in[0,6\ell]$. For a strip geometry, the theory includes gapless edge modes (red). The dotted line highlights the cluster $C(\ell,2)$ used in DMRG simulations. 
}
\end{center}
\end{figure} 

We demonstrate the power of this approach for a honeycomb lattice of spin-1/2 Heisenberg chains linked together by three-spin 
junction couplings, see Fig.~\ref{fig1}. A single Y junction of spin-1/2 Heisenberg chains has been studied in Refs.~\cite{Buccheri2018,Buccheri2019}.  The boundary conditions at this junction can be controlled by tuning a three-spin interaction $J_\chi$, see Fig.~\ref{fig1a}. Remarkably, for a special value $J_\chi=J_\chi^c$, one finds an ideal \emph{chiral fixed point} where incoming spin currents are perfectly 
rerouted to the next chain in rotation, without any backscattering. For $\ell\to \infty$, the chiral point is unstable as it corresponds to a 
BCFT critical point \cite{Buccheri2018,Buccheri2019}.  We here show  from density matrix renormalization group (DMRG) simulations that in 
networks of finite-length spin chains, the chiral fixed point remains present. Moreover, it exists even 
for rather short chains with $\ell=8$.  For the honeycomb lattice in Fig.~\ref{fig1b}, the chiral point 
then begets a non-degenerate and \emph{stable CSL} with energy gap $E_1\propto 1/\ell$. 
For spin-$1/2$ chains, the resulting phase is precisely the Kalmeyer-Laughlin CSL. From our construction, it is straightforward to
establish a quantized spin Hall conductance and the existence of localized spinons with semionic statistics as elementary excitations.   
While we here focus on the Kalmeyer-Laughlin CSL as proof-of-principle example, our construction can readily be generalized to treat non-Abelian CSLs from  higher-$S$ spin chains \cite{Huang2016}. In addition, the case of gapless CSLs can be accessed by using a staggered  chirality (see also Ref.~\cite{Bauer2019}), and one can also describe three-dimensional CSLs, e.g., on a hyperhoneycomb network \cite{OBrien2016}.  In addition,
we anticipate that by allowing for SU($n>2$) spin rotation symmetry, for 
 chiral junctions of more than three chains, and/or by including the effects of a magnetic field, 
 the physics of all CSLs described by the projective symmetry group classification \cite{Bieri2016} will become 
 accessible.  Apart from this  conceptual breakthrough, our chiral-junction network construction may also guide experimental efforts towards engineering synthetic CSL materials. 

\emph{2D network at chiral fixed point.---}We begin with the Hamiltonian for a 2D honeycomb network of spin-1/2 Heisenberg chains, see Fig.~\ref{fig1b}, where
length-$\ell$ chains are coupled by a three-spin boundary interaction $J_\chi$,
\be
H=J\sum_c\sum_{\langle ij\rangle\in  c}\mb S_i\cdot \mb S_j+J_\chi\sum_b\sum_{ijk\in b}\mb S_i\cdot(\mb S_j\times\mb S_k).\label{model}
\ee 
Here $J>0$ is the exchange coupling between nearest-neighbor spin operators $\mb S_i$ and $\mb S_j$ within the same chain $c$. The coupling $J_\chi$ breaks time-reversal symmetry and induces a scalar spin chirality \cite{Wen1989} at the boundary triangles ($b$), see Fig.~\ref{fig1a}. It can be generated in Mott insulators by using circularly polarized light \cite{Claassen2017}. The  model \eqref{model} could be realized in cold atom arrays \cite{Endres2016}, atomic   chains on insulating surfaces \cite{Choi2019}, or in superconducting circuits \cite{Wang2019}.   
 We choose $J_\chi>0$ and order the spins in the triple product such that the triangles $i\to j\to k$ are oriented counterclockwise. 
 The model \eqref{model} has only the dimensionless ratio $g=J_\chi/J$ and the chain length $\ell$ as free parameters. We focus on the case of even $\ell$ where the total spin of each chain is integer. While our field theory below applies for $\ell\gg 1$, we note that  $\ell=2$ corresponds to a star lattice  \cite{Yang2010,Jahromi2018}.
 
 In the large-$\ell$ continuum limit, non-Abelian bosonization expresses the low-energy bulk excitations of the spin chains in terms of SU(2)$_1$ Wess-Zumino-Witten models \cite{Affleck1987,Gogolin1998}.  We associate with each chain a pair of chiral spin currents, $\mb J_{\nu\alpha}(\mb V,x)$, where $\nu=+,-=L,R$ refers to incoming or outgoing modes at the $x=0$ boundary of chain $\alpha$, respectively.
 The 2D vector $\mb V$ specifies the location of junctions corresponding to down-pointing triangles, and  
 $x\in [0,\ell]$ is the 1D coordinate measured from $\mb V$ along a given chain, see~Fig.~\ref{fig1}.
These chiral currents can be represented by chiral boson fields $\varphi_{\nu\alpha}(\mb V,x)$ \cite{Affleck1987,Gogolin1998},
\be \label{bosonization}
J_{\nu\alpha}^z =\frac{\nu}{2\sqrt{\pi}}\partial_x \varphi_{\nu\alpha},\quad J_{\nu\alpha}^\pm =
J^x_{\nu\alpha}\pm iJ^y_{\nu\alpha}= \frac{1}{2\pi}  e^{\pm i2\sqrt{\pi}\varphi_{\nu\alpha}} .
\ee
The physics of the interacting spin network is then encoded by boundary conditions at the Y junctions. 

For a single Y junction with $\ell\gg 1$, the low-energy physics is governed by a chiral fixed point, cf.~Ref.~\cite{Oshikawa2006}, for a critical value $g=g_c(\ell)$ with $g_c(\infty)\approx 3.4$ \cite{Buccheri2018,Buccheri2019}.   Right at the fixed point, chiral spin currents are related by 
a chiral boundary condition.  For the 2D network with ideal chiral junctions, $g=g_c(\ell)$, we thus have the 
boundary conditions ($\alpha=\alpha\mod 3$)
 \be
 \mb J_{R,\alpha}(\mb V,0)=\mb J_{L,\alpha-1}(\mb V,0)\label{bc1}
 \ee 
for the down-pointing triangles in Fig.~\ref{fig1b}. 
 At the up-pointing triangles, the corresponding conditions are given by \cite{foot1}
 \be\label{bc2}
 \mb J_{R,\alpha}(\mb V,\ell)=\mb J_{L,\alpha+1}(\mb V+\boldsymbol\delta_{\alpha-1},\ell),
 \ee
 see also Refs.~\cite{Rahmani2010,Buccheri2019}.
 Normal modes then correspond to chiral boson fields, $\phi(\mb R,\xi)$, circulating in loops around each hexagon.
 We label the hexagon center positions by the vector $\mb R$,
see Fig.~\ref{fig1b}, and use the 1D coordinate $\xi\equiv\xi\mod 6\ell$ along the loop.
 The above picture is reminiscent of the Chalker-Coddington model  
\cite{Chalker1988} and (when neglecting RG-irrelevant operators) becomes asymptotically exact at the chiral fixed point.

 \begin{figure}[t]
\begin{center}
\subfloat[\label{fig2a}]{\includegraphics[width=0.35\columnwidth]{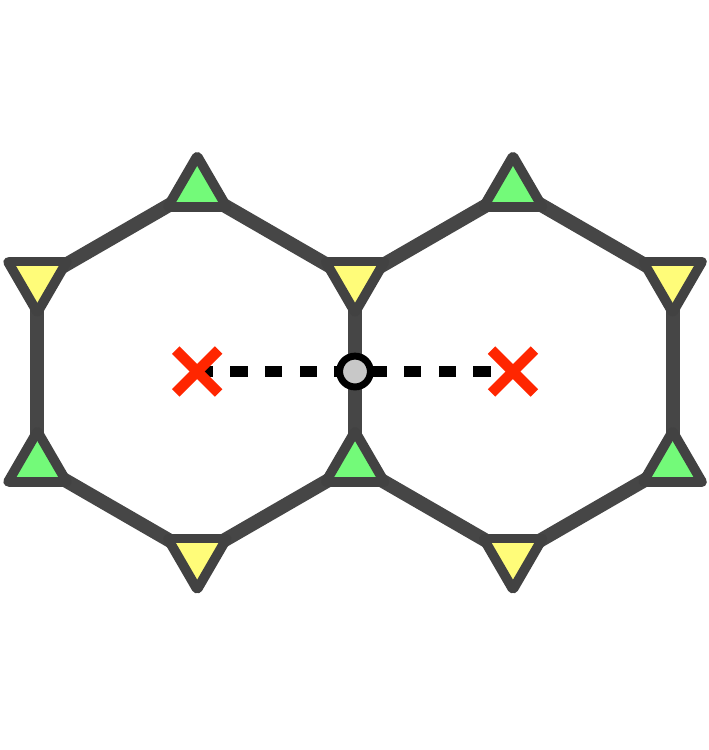}}\qquad
\subfloat[\label{fig2b}]{\includegraphics[width=0.5\columnwidth]{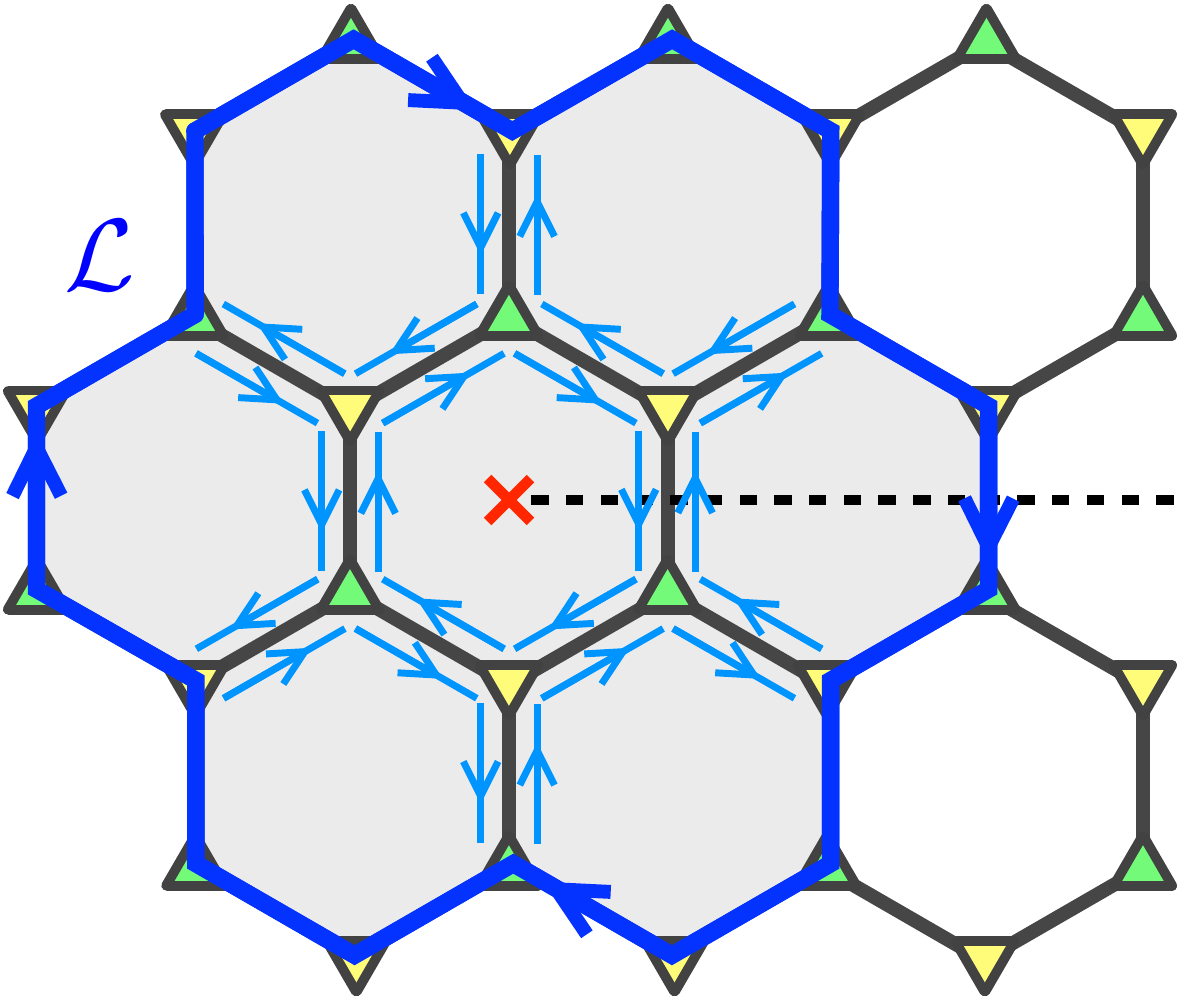}
}
\caption{\label{fig2}Spinons in the CSL phase. (a) A local operator acting on a single site of a given chain $c$ (gray circle) excites the zero modes of two neighboring hexagons. These excitations with $\Delta N(\mb R)=\pm 1/2$ (red crosses) correspond to spinons. The dashed line connecting two spinons defines an open string $w$. (b) The operator transporting a spinon around a loop $\mc L$  (purple line) produces a $\pi$ phase shift if a spinon is present inside $\mc L$. }
\end{center}
\end{figure}

\emph{Gapped spectrum.---}Local operators in general involve chiral bosons belonging to two neighboring hexagons. For instance, the staggered part of the  spin operator $\mb S_j=\mb S_\alpha(\mb V,x)$ can be written as \cite{Gogolin1998}
\bea
S_\alpha^\pm(\mb V,x) &\sim& (-1)^x \exp[\pm i\sqrt\pi(\phi(\mb R',\xi')+\phi(\mb R,\xi))],\nonumber\\
S_\alpha^z(\mb V,x) &\sim &(-1)^x\sin[\sqrt{\pi}(\phi(\mb R',\xi')-\phi(\mb R,\xi))],\label{localops}
\eea
where $(\mb R,\xi)$ and $(\mb R',\xi')$, with $\mb R'=\mb R+\boldsymbol\delta_\alpha$ and $\xi'=2(\alpha-1)\ell-\xi$, refer to the respective  center and  
1D coordinates of the neighboring hexagons.
 A standard mode expansion expresses $\phi(\mb R,\xi)$ in terms of canonically conjugate zero-mode operators $\phi_0(\mb R)$ and $Q(\mb R)$ and
boson annihilation operators $a_n(\mb R)$ for finite momentum $q_n=\pi n/(3\ell)$ with integer $n>0$ 
\cite{Note1}.
Invariance of the local operators \eqref{localops} under $\xi\mapsto \xi+6\ell$ quantizes the eigenvalues of 
$ Q(\mb R)$  as $2\sqrt\pi N(\mb R)$, where $N(\mb R)$ and $N(\mb R')$ for the two hexagons in Eq.~(\ref{localops}) must be both integer or both half-integer.  Using Eq.~\eqref{bosonization}, this selection rule ensures that   
$S^z_{\textrm{tot}} = \frac1{2\sqrt\pi} \sum_{\mb R}\int_0^{6\ell} d\xi \, \partial_\xi\phi=\sum_{\mb R}N(\mb R)$
is integer for all  physical states. We also observe that $N(\mb R)$ determines the local magnetization associated with the chiral boson for the hexagon  at $\mb R$. 
The effective low-energy Hamiltonian at the chiral fixed point then has the form  
\be
H_{\textrm{c}} =\sum_{\mb R}\left(\frac{v}{12\ell}Q^2(\mb R)+\sum_{n>0}\frac{ \pi n v}{3\ell}a_n^\dagger(\mb R)a_{n}^{\phantom\dagger}(\mb R)\right),\label{Hchiral}
\ee
where $v=\pi J/2$ is the spin velocity \cite{Gogolin1998}. The ground state is the vacuum, $Q(\mb R)|0\rangle=a_n(\mb R)|0\rangle=0$ for all $\mb R$ and $n$.  The first excited state is highly degenerate and corresponds to changing the zero-mode eigenvalues of two hexagons by $\Delta N(\mb R)=\pm1/2$, with    energy $E_1=\pi v/(6\ell)$. We refer to the elementary spin-1/2 excitation in a hexagon as \emph{spinon}. Although local operators create spinons in neighboring hexagons, see Fig.~\ref{fig2a}, they can be separated by arbitrary distances without energy cost. To see this, consider the string operator 
$\mc S(w)=\prod_{c}e^{i\sqrt{\pi}\Phi_0(c)}$ with  $\Phi_0(c)=\phi_0(\mb R')-\phi_0(\mb R)$,
which acts on the zero modes of all hexagons sharing chains $c$ crossed by the open string  $w$, see Fig.~\ref{fig2}. 
Using $[\phi_0(\mb R),Q(\mb R')]=i\delta_{\mb R\mb R'}$, one readily finds that the state
$\mc S(w)|0\rangle$ represents a two-spinon excitation with energy $E_1$, where
spinons are localized at the endpoints of the (arbitrarily long) string $w$.

\emph{Topological properties.---}We next show that the spinons defined above are \emph{semions}, a hallmark property of the Kalmeyer-Laughlin CSL \cite{Kalmeyer1987}. 
Our argument is similar to the proof for semionic statistics in the toric code \cite{Kitaev2003}. In the continuum limit, 
the operator $U=e^{i\sqrt\pi\int d\xi\,\partial_\xi\phi(\mb R,\xi)}$ transports a spinon along the chain direction  
\cite{Gorohovsky2015}. We then combine the operators for different chains to an operator $U_{\mc L}$ that takes a 
spinon around a closed path $\mc L$, see Fig.~\ref{fig2b}.  Next we note that
for every chain $c$ between neighboring hexagons $\mb R$ and $\mb R'$, we have $I_c=e^{2\pi i \sum_{j\in c}\,S_j^z}=\mathtt{1}$ because even-$\ell$ chains have integer spin. 
We thus can  multiply $U_{\mc L}$ by  
$I_c \sim e^{i\sqrt\pi\int d\xi\,\partial_\xi\phi(\mb R,\xi)}e^{i\sqrt\pi\int d\xi'\,\partial_{\xi'}\phi(\mb R',\xi')}$
for all chains in the region $\mc A$  bounded by $\mc L$, i.e., the shaded area in Fig.~\ref{fig2b}. The inner loops are thereby completed and the Stokes theorem gives
\be
U_{\mc L}=e^{i\sqrt\pi\sum_{\mb R\in \mc A}\int_0^{6\ell}d\xi\,\partial_\xi\phi(\mb R,\xi)}=e^{i2\pi\sum_{\mb R\in \mc A}N(\mb R)}.
\ee
For an odd number of spinons inside $\mc L$, we have $U_{\mc L}=-1$ as expected for semionic statistics. We note that in contrast to parallel-chain constructions \cite{Gorohovsky2015}, our analysis of fractional statistics does not hinge on  semiclassical approximations of the effective field theory.

Another important property of the CSL is its quantized spin Hall conductance. In a finite-size network at the chiral fixed point, we must have  
gapless edge modes decoupled from bulk modes, see Fig.~\ref{fig1b}. In a strip geometry of width $W\gg \ell$, the two edge modes along the strip direction 
can be treated as (spatially separated) left- and right-moving
chiral boson fields $\phi_{\nu=L,R}(\xi)$ with the edge Hamiltonian $H_{\textrm{edge}}=\sum_{\nu}\frac{v}2\int_{-\infty}^{\infty}d\xi\,(\partial_\xi\phi_\nu)^2$. 
Applying opposite magnetic fields at the two  edges by adding the terms $\delta H_{\pm}=\pm \frac{h}2\sum_{j\in\textrm{edge}}S_j^z$, one imposes a transverse spin voltage $h\ll E_1$. Using Eq.~\eqref{bosonization}, we 
obtain the longitudinal spin current 
\be
J_s=\frac{v}{2\sqrt\pi}\left\langle \partial_\xi\phi_L+\partial_\xi\phi_R\right\rangle=\frac{h}{4\pi},
\ee 
which indeed yields the quantized spin Hall conductance $\sigma^s_{xy}=1/2$ in units of the spin conductance quantum  $\frac1{2\pi}$ (with $\hbar=1$) \cite{YaoLee2011,Kumar2015}. 

\begin{figure}[t]
\begin{center}
{\includegraphics[width=.9\columnwidth]{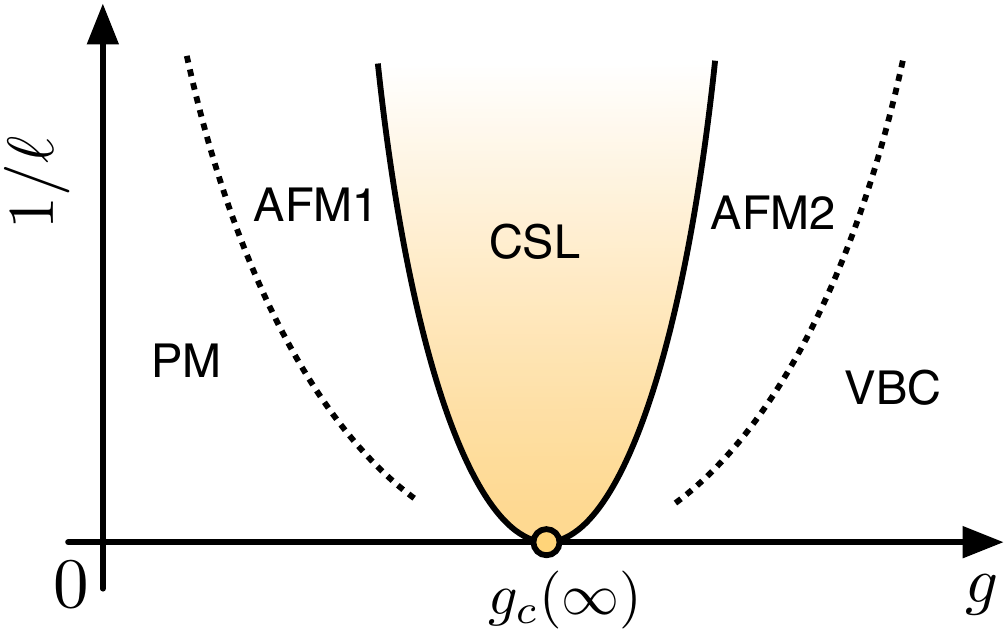}}
\caption{\label{fig3}
Schematic phase diagram for the 2D network \eqref{model}, where the shaded region indicates a stable CSL phase.
In addition to quantum paramagnetic (PM) and valence bond crystal (VBC) phases, we also expect antiferromagnetically ordered phases (AFM1 and AFM2). 
}
\end{center}
\end{figure}

\emph{Phase diagram.---}Away from the chiral point, the field theory contains a relevant perturbation due to backscattering at the junctions 
\cite{Buccheri2018,Buccheri2019}. For the 2D network, this term is given by
\be
H_{\textrm{bs}}=\lambda \sum_{\mb R,\alpha,r} \cos[\sqrt\pi(\phi(\mb R+\boldsymbol\delta_\alpha,\xi'_{\alpha r})-\phi(\mb R,\xi_{\alpha r}))],\label{backscattering}
\ee
where $\lambda\propto g-g_c$ for $|g-g_c|\ll 1$.  
The sum over $\mb R$ and $\alpha=1,2,3$ counts each hexagon pair once, where backscattering can occur at the two shared Y junctions (labeled by $r=1,2$) and 
the 1D coordinates at the respective junction locations are $\xi_{\alpha r}$ and $\xi'_{\alpha r}$.  Right at the chiral fixed point $g=g_c(\ell)$, we have $\lambda=0$. 
Importantly, Eq.~(\ref{backscattering}) allows for quasiparticle scattering between hexagons such that spinons are no longer localized for $\lambda\neq 0$.  
Using first-order degenerate perturbation theory to compute the matrix elements of $H_{\textrm{bs}}$ between 
two-spinon states $\mc S(w)|0\rangle$, we obtain   
a tight-binding model of spinons on the triangular hexagon lattice with 
hopping parameter $t_{\textrm{eff}}\propto \lambda\ell^{-1/2}$.
As a result, the degeneracy of the first excited state is lifted.
A phase transition occurs once the spinon bandwidth closes the energy gap, i.e., for $|t_{\textrm{eff}}|\sim E_1$.  
Using $\lambda\propto g-g_c$, the gapped CSL phase is thus stable for 
$[g-g_c(\ell)]^2 <  \frac{c_0}{\ell}$ with $c_0={\cal O}(1)$.
We sketch the phase diagram in Fig.~\ref{fig3}, where the paramagnetic (PM) and valence bond crystal (VBC) regions are briefly discussed below. 
Since spinons condense at distinct wave vectors for $\lambda<0$ and $\lambda>0$, different magnetically ordered regions (AFM1 and AFM2) 
are also expected, see Fig.~\ref{fig3}. However, a detailed discussion will be given elsewhere.

\begin{figure}[t]
\begin{center}
{\includegraphics[width=\columnwidth]{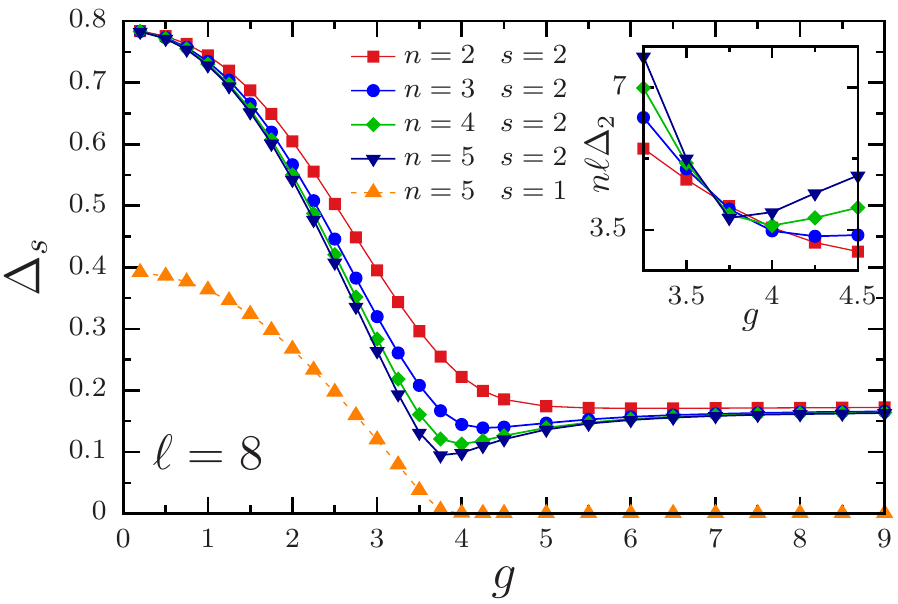}}
\caption{\label{fig4}
DMRG results for the energy gaps $\Delta_{s=1,2}$ to the first excited state with total spin $S^z_{\rm tot}=s$ vs the coupling  
$g$ for $C(\ell=8,n)$ clusters. The singlet-triplet gap $\Delta_1$ drops to zero for $g\gtrsim 3.7$. The gap $\Delta_2$ has a minimum with 
$\Delta_2^{\rm (min)}\propto 1/(n\ell)$ at $g=g_c(\ell)$. Inset: Magnified view of $n\ell \Delta_2$ vs $g$.}
\end{center}
\end{figure}

\emph{DMRG results.---}We have implemented the DMRG method \cite{White1992} for Eq.~\eqref{model} on 
tree-like clusters $C(\ell,n)$  containing $n$  unit cells along one direction but only one cell along the other, see 
Fig.~\ref{fig1b} for $C(\ell,2)$, keeping up to 1000 states per DMRG block. 
For a similar comb geometry implemented in a tensor-network based algorithm, see  Ref.~\cite{Chepiga2019}.  Since the clusters $C(\ell,n)$ contain no loops, our DMRG results cannot provide direct CSL evidence. Nonetheless, they (i) reveal the chiral fixed point for a network
with many Y junctions where (ii) intermediate values of $\ell$ suffice for realizing the chiral point (here $\ell=8$). In addition, (iii) the DMRG results support the phase diagram  in Fig.~\ref{fig3}. 

Figure \ref{fig4} shows the 
energy gap $\Delta_{s=1,2}$ to the first excited state with total magnetization $S^z_{\textrm{tot}}=s$. 
For $g\ll 1$, both $\Delta_1$ and $\Delta_2$ are nonzero and almost independent of $n$. The ground state is then adiabatically  connected to a product of singlets on decoupled chains which is the exact ground state for $g\to 0$. 
On the 2D network, this state corresponds to the PM phase with singlet-triplet gap $\Delta_1\propto 1/\ell$. 
For $g\gg1 $,  $\Delta_1$ decreases exponentially with $n$ while $\Delta_2$ remains finite, 
implying a fourfold degenerate ground state. In fact, for $g\to\infty$, an effective spin-$1/2$ operator 
emerges from the three strongly coupled spins at each junction, with exchange coupling $J'=J/3$ 
to the boundary spins of the remaining length-$(\ell-2)$ chains \cite{Buccheri2018}. 
We find a VBC phase with a pattern of one strong and two weak bonds to emergent spins, see Ref.~\footnote{See the accompanying Supplementary Material, where we present additional DMRG results.}, where each cluster boundary hosts an effective spin-1/2 and
the chiral point separates phases without or with end spins.  
Right at the chiral point, a chiral edge mode propagates around the entire system. 
We then expect the gaps to vanish as $\Delta_s\sim  1/n$ for $n\to \infty$. 
The inset of Fig.~\ref{fig4} confirms this behavior for $g \approx 3.7$.  We have thus identified the chiral point in this geometry, $g_c(\ell=8)\approx 3.7$. 
 
\emph{Conclusions.---}Our CSL construction employs chiral junctions of multiple spin chains as basic building blocks. 
Like the thin torus limit of the quantum Hall effect \cite{Bergholtz2005}, this approach expresses the essential physics in 
more feasible geometries. Indeed, since the gap scales as $E_1\propto 1/\ell$, 
the CSL phase can already be accessed for networks with rather short $\ell$.  
Our approach paves the way to a systematic study of many other CSL phases through the general 
connection between CSLs and chiral critical points in BCFT.  

\begin{acknowledgments}
We thank F. Buccheri, C. Chamon and S. Manmana for helpful discussions, and the High-Performance Computing Center (NPAD) at UFRN for providing computational resources. G. F. acknowledges support from Capes. Research at   IIP-UFRN is supported by Brazilian ministries MEC and MCTIC. We acknowledge funding by the Deutsche Forschungsgemeinschaft (DFG, German Research Foundation),
Projektnummer 277101999 - TRR 183 (project C04), and by the Humboldt foundation under the Bessel award program.
\end{acknowledgments}

\bibliographystyle{aipnum4-1}

%

\onecolumngrid

\appendix
\section{Supplementary Material to ``Spin chain network construction of chiral spin liquids''}

\subsection{Bosonization details}

The mode expansion for the boson fields $\phi(\mb R,\xi)$ employed in the main text is given by
\be
\phi(\mb R,\xi)=\phi_0(\mb R)+\frac{Q(\mb R)\xi}{6\ell}+\sum_{n>0}\frac{a_n(\mb R)e^{\frac{i\pi n \xi}{3\ell}}+\textrm{h.c.}}{\sqrt{2\pi n}},
\ee
where $a_n(\mb R)$ are boson annihilation operators associated with momentum $q_n=\pi n/(3\ell)$, and $\phi_0(\mb R)$ and $Q(\mb R)$  are zero-mode operators with $[\varphi_0(\mb R),Q(\mb R')]=i\delta_{\mb R\mb R'}$.\\

\begin{figure}[b]
\begin{center}
\subfloat[\label{figSMa}]{\includegraphics[width=.4\columnwidth]{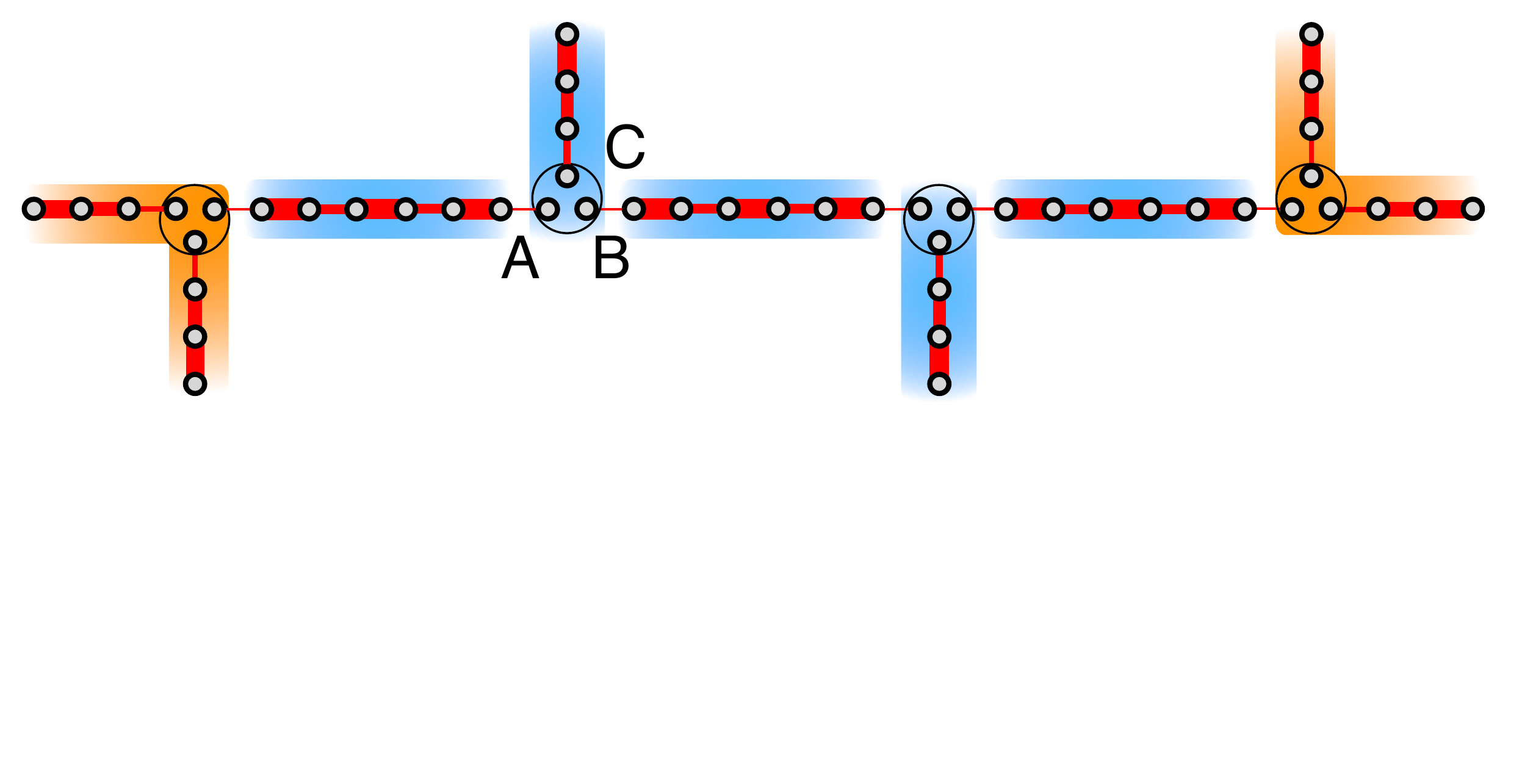}}\qquad 
\subfloat[\label{figSMb}]{\includegraphics[width=.5\columnwidth]{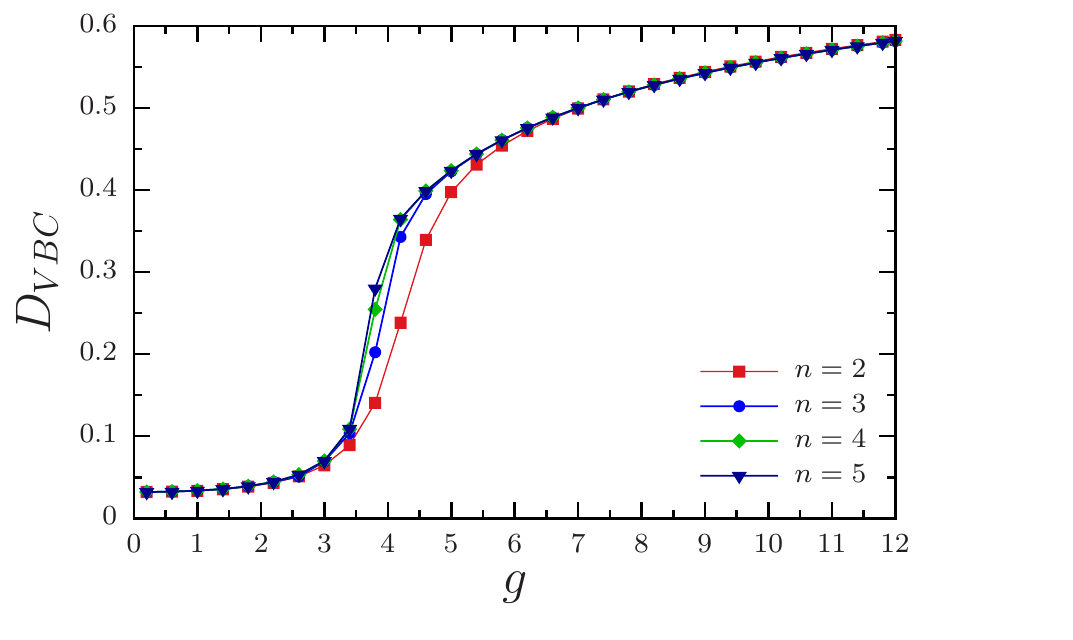}}
\caption{
Dimerization in clusters of spin chains. (a) Dimerization pattern in the $C(8,2)$ cluster with $g=20$. The thickness of each red line is proportional to the magnitude of the bond. The emergent spin-1/2 degrees of freedom at each Y junction are indicated by  circles.
The bonds between this spin and the boundary spins of the residual chains are labeled by A,B,C and exhibit the VBC pattern 
described in the text.
(b) DMRG results for the VBC order parameter $D_{\rm VBC}$, see Eq.~\eqref{VBC}, vs the 
chiral interaction parameter $g$ for $C(8,n)$ clusters. 
}
\end{center}
\end{figure}

\subsection{Additional DMRG results}

Our DMRG simulations for the clusters $C(\ell,n)$ provide evidence for the VBC phase 
when the chiral coupling $g=J_\chi/J$ is large, $g\gg1$.  We describe these results here.
For $g\gg 1$, the three spins at a Y junction are strongly coupled.  At low energy scales, they 
form an emergent spin-1/2 degree of freedom that then couples by a weaker exchange coupling $J'=J/3$
to the boundary spins of the residual spin chains.  These three bonds are labeled by A,B,C in the figure.
In the VBC phase, one expects a characteristic pattern with one strong and two weak bonds at each junction, where
 the bond strength refers to the value of $|\langle \mb S_i\cdot \mb S_j\rangle|$.
This pattern is quantified by the VBC order parameter,
\be\label{VBC}
D_{\textrm{VBC}}=1-\frac{\left\langle\mb S_i\cdot\mb S_j\right\rangle_{\textrm{A}}+\left\langle\mb S_i\cdot\mb S_j\right\rangle_{\textrm{B}}}{2\left\langle\mb S_i\cdot\mb S_j\right\rangle_{\textrm{C}}}.
\ee
While the VBC pattern breaks the rotational symmetry of the 2D network, it does not break any symmetry for the cluster geometry. 
As shown in the figure, our DMRG results indicate that for $\ell=8$, the order parameter $D_{\textrm{VBC}}$ is a smooth function of 
$g$ which significantly increases for $g\gg 1$.  
A closer analysis of the bond pattern reveals that the ground state 
can be pictured as a product of singlets in the bulk (blue regions in the inset) and two outer blocks 
containing $\ell+1$ spins (orange regions).  In total, these outer regions realize an effective spin-1/2 degree of freedom 
localized at the ends of the cluster. This observation explains the fourfold degeneracy found for $n\to\infty$ in our DMRG results.


\begin{thebibliography}{53}%
\makeatletter
\providecommand \@ifxundefined [1]{%
 \@ifx{#1\undefined}
}%
\providecommand \@ifnum [1]{%
 \ifnum #1\expandafter \@firstoftwo
 \else \expandafter \@secondoftwo
 \fi
}%
\providecommand \@ifx [1]{%
 \ifx #1\expandafter \@firstoftwo
 \else \expandafter \@secondoftwo
 \fi
}%
\providecommand \natexlab [1]{#1}%
\providecommand \enquote  [1]{``#1''}%
\providecommand \bibnamefont  [1]{#1}%
\providecommand \bibfnamefont [1]{#1}%
\providecommand \citenamefont [1]{#1}%
\providecommand \href@noop [0]{\@secondoftwo}%
\providecommand \href [0]{\begingroup \@sanitize@url \@href}%
\providecommand \@href[1]{\@@startlink{#1}\@@href}%
\providecommand \@@href[1]{\endgroup#1\@@endlink}%
\providecommand \@sanitize@url [0]{\catcode `\\12\catcode `\$12\catcode
  `\&12\catcode `\#12\catcode `\^12\catcode `\_12\catcode `\%12\relax}%
\providecommand \@@startlink[1]{}%
\providecommand \@@endlink[0]{}%
\providecommand \url  [0]{\begingroup\@sanitize@url \@url }%
\providecommand \@url [1]{\endgroup\@href {#1}{\urlprefix }}%
\providecommand \urlprefix  [0]{URL }%
\providecommand \Eprint [0]{\href }%
\providecommand \doibase [0]{http://dx.doi.org/}%
\providecommand \selectlanguage [0]{\@gobble}%
\providecommand \bibinfo  [0]{\@secondoftwo}%
\providecommand \bibfield  [0]{\@secondoftwo}%
\providecommand \translation [1]{[#1]}%
\providecommand \BibitemOpen [0]{}%
\providecommand \bibitemStop [0]{}%
\providecommand \bibitemNoStop [0]{.\EOS\space}%
\providecommand \EOS [0]{\spacefactor3000\relax}%
\providecommand \BibitemShut  [1]{\csname bibitem#1\endcsname}%
\let\auto@bib@innerbib\@empty
\bibitem [{\citenamefont {Wen}(2017)}]{Wen2017}%
  \BibitemOpen
  \bibfield  {author} {\bibinfo {author} {\bibfnamefont {X.-G.}\ \bibnamefont
  {Wen}},\ }\href {https://link.aps.org/doi/10.1103/RevModPhys.89.041004}
  {\bibfield  {journal} {\bibinfo  {journal} {Rev. Mod. Phys.}\ }\textbf
  {\bibinfo {volume} {89}},\ \bibinfo {pages} {041004} (\bibinfo {year}
  {2017})}\BibitemShut {NoStop}%
\bibitem [{\citenamefont {Savary}\ and\ \citenamefont
  {Balents}(2017)}]{Savary2016}%
  \BibitemOpen
  \bibfield  {author} {\bibinfo {author} {\bibfnamefont {L.}~\bibnamefont
  {Savary}}\ and\ \bibinfo {author} {\bibfnamefont {L.}~\bibnamefont
  {Balents}},\ }\href {http://stacks.iop.org/0034-4885/80/i=1/a=016502}
  {\bibfield  {journal} {\bibinfo  {journal} {Rep. Prog. Phys.}\ }\textbf
  {\bibinfo {volume} {80}},\ \bibinfo {pages} {016502} (\bibinfo {year}
  {2017})}\BibitemShut {NoStop}%
\bibitem [{\citenamefont {Knolle}\ and\ \citenamefont
  {Moessner}(2019)}]{Knolle2019}%
  \BibitemOpen
  \bibfield  {author} {\bibinfo {author} {\bibfnamefont {J.}~\bibnamefont
  {Knolle}}\ and\ \bibinfo {author} {\bibfnamefont {R.}~\bibnamefont
  {Moessner}},\ }\href
  {https://doi.org/10.1146/annurev-conmatphys-031218-013401} {\bibfield
  {journal} {\bibinfo  {journal} {Annu. Rev. Condens. Matter Phys.}\ }\textbf
  {\bibinfo {volume} {10}},\ \bibinfo {pages} {451} (\bibinfo {year}
  {2019})}\BibitemShut {NoStop}%
\bibitem [{\citenamefont {Kalmeyer}\ and\ \citenamefont
  {Laughlin}(1987)}]{Kalmeyer1987}%
  \BibitemOpen
  \bibfield  {author} {\bibinfo {author} {\bibfnamefont {V.}~\bibnamefont
  {Kalmeyer}}\ and\ \bibinfo {author} {\bibfnamefont {R.~B.}\ \bibnamefont
  {Laughlin}},\ }\href {\doibase 10.1103/PhysRevLett.59.2095} {\bibfield
  {journal} {\bibinfo  {journal} {Phys. Rev. Lett.}\ }\textbf {\bibinfo
  {volume} {59}},\ \bibinfo {pages} {2095} (\bibinfo {year}
  {1987})}\BibitemShut {NoStop}%
\bibitem [{\citenamefont {Wen}, \citenamefont {Wilczek},\ and\ \citenamefont
  {Zee}(1989)}]{Wen1989}%
  \BibitemOpen
  \bibfield  {author} {\bibinfo {author} {\bibfnamefont {X.~G.}\ \bibnamefont
  {Wen}}, \bibinfo {author} {\bibfnamefont {F.}~\bibnamefont {Wilczek}}, \ and\
  \bibinfo {author} {\bibfnamefont {A.}~\bibnamefont {Zee}},\ }\href
  {https://link.aps.org/doi/10.1103/PhysRevB.39.11413} {\bibfield  {journal}
  {\bibinfo  {journal} {Phys. Rev. B}\ }\textbf {\bibinfo {volume} {39}},\
  \bibinfo {pages} {11413} (\bibinfo {year} {1989})}\BibitemShut {NoStop}%
\bibitem [{\citenamefont {Kitaev}(2006)}]{Kitaev2006}%
  \BibitemOpen
  \bibfield  {author} {\bibinfo {author} {\bibfnamefont {A.}~\bibnamefont
  {Kitaev}},\ }\href
  {http://www.sciencedirect.com/science/article/pii/S0003491605002381}
  {\bibfield  {journal} {\bibinfo  {journal} {Ann. Phys.}\ }\textbf {\bibinfo
  {volume} {321}},\ \bibinfo {pages} {2} (\bibinfo {year} {2006})}\BibitemShut
  {NoStop}%
\bibitem [{\citenamefont {Kasahara}\ \emph {et~al.}(2018)\citenamefont
  {Kasahara}, \citenamefont {Ohnishi}, \citenamefont {Mizukami}, \citenamefont
  {Tanaka}, \citenamefont {Ma}, \citenamefont {Sugii}, \citenamefont {Kurita},
  \citenamefont {Tanaka}, \citenamefont {Nasu}, \citenamefont {Motome},
  \citenamefont {Shibauchi},\ and\ \citenamefont {Matsuda}}]{Kasahara2018}%
  \BibitemOpen
  \bibfield  {author} {\bibinfo {author} {\bibfnamefont {Y.}~\bibnamefont
  {Kasahara}}, \bibinfo {author} {\bibfnamefont {T.}~\bibnamefont {Ohnishi}},
  \bibinfo {author} {\bibfnamefont {Y.}~\bibnamefont {Mizukami}}, \bibinfo
  {author} {\bibfnamefont {O.}~\bibnamefont {Tanaka}}, \bibinfo {author}
  {\bibfnamefont {S.}~\bibnamefont {Ma}}, \bibinfo {author} {\bibfnamefont
  {K.}~\bibnamefont {Sugii}}, \bibinfo {author} {\bibfnamefont
  {N.}~\bibnamefont {Kurita}}, \bibinfo {author} {\bibfnamefont
  {H.}~\bibnamefont {Tanaka}}, \bibinfo {author} {\bibfnamefont
  {J.}~\bibnamefont {Nasu}}, \bibinfo {author} {\bibfnamefont {Y.}~\bibnamefont
  {Motome}}, \bibinfo {author} {\bibfnamefont {T.}~\bibnamefont {Shibauchi}}, \
  and\ \bibinfo {author} {\bibfnamefont {Y.}~\bibnamefont {Matsuda}},\ }\href
  {https://doi.org/10.1038/s41586-018-0274-0} {\bibfield  {journal} {\bibinfo
  {journal} {Nature}\ }\textbf {\bibinfo {volume} {559}},\ \bibinfo {pages}
  {227} (\bibinfo {year} {2018})}\BibitemShut {NoStop}%
\bibitem [{\citenamefont {Greiter}\ and\ \citenamefont
  {Thomale}(2009)}]{Greiter2009}%
  \BibitemOpen
  \bibfield  {author} {\bibinfo {author} {\bibfnamefont {M.}~\bibnamefont
  {Greiter}}\ and\ \bibinfo {author} {\bibfnamefont {R.}~\bibnamefont
  {Thomale}},\ }\href {\doibase 10.1103/PhysRevLett.102.207203} {\bibfield
  {journal} {\bibinfo  {journal} {Phys. Rev. Lett.}\ }\textbf {\bibinfo
  {volume} {102}},\ \bibinfo {pages} {207203} (\bibinfo {year}
  {2009})}\BibitemShut {NoStop}%
\bibitem [{\citenamefont {Chua}, \citenamefont {Yao},\ and\ \citenamefont
  {Fiete}(2011)}]{Chua2011}%
  \BibitemOpen
  \bibfield  {author} {\bibinfo {author} {\bibfnamefont {V.}~\bibnamefont
  {Chua}}, \bibinfo {author} {\bibfnamefont {H.}~\bibnamefont {Yao}}, \ and\
  \bibinfo {author} {\bibfnamefont {G.~A.}\ \bibnamefont {Fiete}},\ }\href
  {http://link.aps.org/doi/10.1103/PhysRevB.83.180412} {\bibfield  {journal}
  {\bibinfo  {journal} {Phys. Rev. B}\ }\textbf {\bibinfo {volume} {83}},\
  \bibinfo {pages} {180412(R)} (\bibinfo {year} {2011})}\BibitemShut {NoStop}%
\bibitem [{\citenamefont {Yao}\ and\ \citenamefont {Lee}(2011)}]{YaoLee2011}%
  \BibitemOpen
  \bibfield  {author} {\bibinfo {author} {\bibfnamefont {H.}~\bibnamefont
  {Yao}}\ and\ \bibinfo {author} {\bibfnamefont {D.-H.}\ \bibnamefont {Lee}},\
  }\href {https://link.aps.org/doi/10.1103/PhysRevLett.107.087205} {\bibfield
  {journal} {\bibinfo  {journal} {Phys. Rev. Lett.}\ }\textbf {\bibinfo
  {volume} {107}},\ \bibinfo {pages} {087205} (\bibinfo {year}
  {2011})}\BibitemShut {NoStop}%
\bibitem [{\citenamefont {Bauer}\ \emph {et~al.}(2014)\citenamefont {Bauer},
  \citenamefont {Cincio}, \citenamefont {Keller}, \citenamefont {Dolfi},
  \citenamefont {Vidal}, \citenamefont {Trebst},\ and\ \citenamefont
  {Ludwig}}]{Bauer2014}%
  \BibitemOpen
  \bibfield  {author} {\bibinfo {author} {\bibfnamefont {B.}~\bibnamefont
  {Bauer}}, \bibinfo {author} {\bibfnamefont {L.}~\bibnamefont {Cincio}},
  \bibinfo {author} {\bibfnamefont {B.~P.}\ \bibnamefont {Keller}}, \bibinfo
  {author} {\bibfnamefont {M.}~\bibnamefont {Dolfi}}, \bibinfo {author}
  {\bibfnamefont {G.}~\bibnamefont {Vidal}}, \bibinfo {author} {\bibfnamefont
  {S.}~\bibnamefont {Trebst}}, \ and\ \bibinfo {author} {\bibfnamefont
  {A.~W.~W.}\ \bibnamefont {Ludwig}},\ }\href
  {http://dx.doi.org/10.1038/ncomms6137} {\bibfield  {journal} {\bibinfo
  {journal} {Nat. Comm.}\ }\textbf {\bibinfo {volume} {5}},\ \bibinfo {pages}
  {5137} (\bibinfo {year} {2014})}\BibitemShut {NoStop}%
\bibitem [{\citenamefont {He}, \citenamefont {Sheng},\ and\ \citenamefont
  {Chen}(2014)}]{He2014}%
  \BibitemOpen
  \bibfield  {author} {\bibinfo {author} {\bibfnamefont {Y.-C.}\ \bibnamefont
  {He}}, \bibinfo {author} {\bibfnamefont {D.~N.}\ \bibnamefont {Sheng}}, \
  and\ \bibinfo {author} {\bibfnamefont {Y.}~\bibnamefont {Chen}},\ }\href
  {\doibase 10.1103/PhysRevLett.112.137202} {\bibfield  {journal} {\bibinfo
  {journal} {Phys. Rev. Lett.}\ }\textbf {\bibinfo {volume} {112}},\ \bibinfo
  {pages} {137202} (\bibinfo {year} {2014})}\BibitemShut {NoStop}%
\bibitem [{\citenamefont {Gorohovsky}, \citenamefont {Pereira},\ and\
  \citenamefont {Sela}(2015)}]{Gorohovsky2015}%
  \BibitemOpen
  \bibfield  {author} {\bibinfo {author} {\bibfnamefont {G.}~\bibnamefont
  {Gorohovsky}}, \bibinfo {author} {\bibfnamefont {R.~G.}\ \bibnamefont
  {Pereira}}, \ and\ \bibinfo {author} {\bibfnamefont {E.}~\bibnamefont
  {Sela}},\ }\href {https://link.aps.org/doi/10.1103/PhysRevB.91.245139}
  {\bibfield  {journal} {\bibinfo  {journal} {Phys. Rev. B}\ }\textbf {\bibinfo
  {volume} {91}},\ \bibinfo {pages} {245139} (\bibinfo {year}
  {2015})}\BibitemShut {NoStop}%
\bibitem [{\citenamefont {Meng}\ \emph {et~al.}(2015)\citenamefont {Meng},
  \citenamefont {Neupert}, \citenamefont {Greiter},\ and\ \citenamefont
  {Thomale}}]{Meng2015}%
  \BibitemOpen
  \bibfield  {author} {\bibinfo {author} {\bibfnamefont {T.}~\bibnamefont
  {Meng}}, \bibinfo {author} {\bibfnamefont {T.}~\bibnamefont {Neupert}},
  \bibinfo {author} {\bibfnamefont {M.}~\bibnamefont {Greiter}}, \ and\
  \bibinfo {author} {\bibfnamefont {R.}~\bibnamefont {Thomale}},\ }\href
  {https://link.aps.org/doi/10.1103/PhysRevB.91.241106} {\bibfield  {journal}
  {\bibinfo  {journal} {Phys. Rev. B}\ }\textbf {\bibinfo {volume} {91}},\
  \bibinfo {pages} {241106(R)} (\bibinfo {year} {2015})}\BibitemShut {NoStop}%
\bibitem [{\citenamefont {Huang}\ \emph {et~al.}(2016)\citenamefont {Huang},
  \citenamefont {Chen}, \citenamefont {Gomes}, \citenamefont {Neupert},
  \citenamefont {Chamon},\ and\ \citenamefont {Mudry}}]{Huang2016}%
  \BibitemOpen
  \bibfield  {author} {\bibinfo {author} {\bibfnamefont {P.-H.}\ \bibnamefont
  {Huang}}, \bibinfo {author} {\bibfnamefont {J.-H.}\ \bibnamefont {Chen}},
  \bibinfo {author} {\bibfnamefont {P.~R.~S.}\ \bibnamefont {Gomes}}, \bibinfo
  {author} {\bibfnamefont {T.}~\bibnamefont {Neupert}}, \bibinfo {author}
  {\bibfnamefont {C.}~\bibnamefont {Chamon}}, \ and\ \bibinfo {author}
  {\bibfnamefont {C.}~\bibnamefont {Mudry}},\ }\href
  {https://link.aps.org/doi/10.1103/PhysRevB.93.205123} {\bibfield  {journal}
  {\bibinfo  {journal} {Phys. Rev. B}\ }\textbf {\bibinfo {volume} {93}},\
  \bibinfo {pages} {205123} (\bibinfo {year} {2016})}\BibitemShut {NoStop}%
\bibitem [{\citenamefont {Lecheminant}\ and\ \citenamefont
  {Tsvelik}(2017)}]{Lecheminant2017}%
  \BibitemOpen
  \bibfield  {author} {\bibinfo {author} {\bibfnamefont {P.}~\bibnamefont
  {Lecheminant}}\ and\ \bibinfo {author} {\bibfnamefont {A.~M.}\ \bibnamefont
  {Tsvelik}},\ }\href {\doibase 10.1103/PhysRevB.95.140406} {\bibfield
  {journal} {\bibinfo  {journal} {Phys. Rev. B}\ }\textbf {\bibinfo {volume}
  {95}},\ \bibinfo {pages} {140406(R)} (\bibinfo {year} {2017})}\BibitemShut
  {NoStop}%
\bibitem [{\citenamefont {Kumar}, \citenamefont {Sun},\ and\ \citenamefont
  {Fradkin}(2015)}]{Kumar2015}%
  \BibitemOpen
  \bibfield  {author} {\bibinfo {author} {\bibfnamefont {K.}~\bibnamefont
  {Kumar}}, \bibinfo {author} {\bibfnamefont {K.}~\bibnamefont {Sun}}, \ and\
  \bibinfo {author} {\bibfnamefont {E.}~\bibnamefont {Fradkin}},\ }\href
  {\doibase 10.1103/PhysRevB.92.094433} {\bibfield  {journal} {\bibinfo
  {journal} {Phys. Rev. B}\ }\textbf {\bibinfo {volume} {92}},\ \bibinfo
  {pages} {094433} (\bibinfo {year} {2015})}\BibitemShut {NoStop}%
\bibitem [{\citenamefont {Sedrakyan}, \citenamefont {Glazman},\ and\
  \citenamefont {Kamenev}(2015)}]{Sedrakyan2015}%
  \BibitemOpen
  \bibfield  {author} {\bibinfo {author} {\bibfnamefont {T.~A.}\ \bibnamefont
  {Sedrakyan}}, \bibinfo {author} {\bibfnamefont {L.~I.}\ \bibnamefont
  {Glazman}}, \ and\ \bibinfo {author} {\bibfnamefont {A.}~\bibnamefont
  {Kamenev}},\ }\href {https://link.aps.org/doi/10.1103/PhysRevLett.114.037203}
  {\bibfield  {journal} {\bibinfo  {journal} {Phys. Rev. Lett.}\ }\textbf
  {\bibinfo {volume} {114}},\ \bibinfo {pages} {037203} (\bibinfo {year}
  {2015})}\BibitemShut {NoStop}%
\bibitem [{\citenamefont {Poilblanc}, \citenamefont {Cirac},\ and\
  \citenamefont {Schuch}(2015)}]{Poilblanc2015}%
  \BibitemOpen
  \bibfield  {author} {\bibinfo {author} {\bibfnamefont {D.}~\bibnamefont
  {Poilblanc}}, \bibinfo {author} {\bibfnamefont {J.~I.}\ \bibnamefont
  {Cirac}}, \ and\ \bibinfo {author} {\bibfnamefont {N.}~\bibnamefont
  {Schuch}},\ }\href {\doibase 10.1103/PhysRevB.91.224431} {\bibfield
  {journal} {\bibinfo  {journal} {Phys. Rev. B}\ }\textbf {\bibinfo {volume}
  {91}},\ \bibinfo {pages} {224431} (\bibinfo {year} {2015})}\BibitemShut
  {NoStop}%
\bibitem [{\citenamefont {Hickey}\ \emph {et~al.}(2016)\citenamefont {Hickey},
  \citenamefont {Cincio}, \citenamefont {Papi\ifmmode~\acute{c}\else
  \'{c}\fi{}},\ and\ \citenamefont {Paramekanti}}]{Hickey2016}%
  \BibitemOpen
  \bibfield  {author} {\bibinfo {author} {\bibfnamefont {C.}~\bibnamefont
  {Hickey}}, \bibinfo {author} {\bibfnamefont {L.}~\bibnamefont {Cincio}},
  \bibinfo {author} {\bibfnamefont {Z.}~\bibnamefont
  {Papi\ifmmode~\acute{c}\else \'{c}\fi{}}}, \ and\ \bibinfo {author}
  {\bibfnamefont {A.}~\bibnamefont {Paramekanti}},\ }\href {\doibase
  10.1103/PhysRevLett.116.137202} {\bibfield  {journal} {\bibinfo  {journal}
  {Phys. Rev. Lett.}\ }\textbf {\bibinfo {volume} {116}},\ \bibinfo {pages}
  {137202} (\bibinfo {year} {2016})}\BibitemShut {NoStop}%
\bibitem [{\citenamefont {Wietek}\ and\ \citenamefont
  {L\"auchli}(2017)}]{Wietek2017}%
  \BibitemOpen
  \bibfield  {author} {\bibinfo {author} {\bibfnamefont {A.}~\bibnamefont
  {Wietek}}\ and\ \bibinfo {author} {\bibfnamefont {A.~M.}\ \bibnamefont
  {L\"auchli}},\ }\href {https://link.aps.org/doi/10.1103/PhysRevB.95.035141}
  {\bibfield  {journal} {\bibinfo  {journal} {Phys. Rev. B}\ }\textbf {\bibinfo
  {volume} {95}},\ \bibinfo {pages} {035141} (\bibinfo {year}
  {2017})}\BibitemShut {NoStop}%
\bibitem [{\citenamefont {Yao}\ \emph {et~al.}(2018)\citenamefont {Yao},
  \citenamefont {Zaletel}, \citenamefont {Stamper-Kurn},\ and\ \citenamefont
  {Vishwanath}}]{Yao2018}%
  \BibitemOpen
  \bibfield  {author} {\bibinfo {author} {\bibfnamefont {N.~Y.}\ \bibnamefont
  {Yao}}, \bibinfo {author} {\bibfnamefont {M.~P.}\ \bibnamefont {Zaletel}},
  \bibinfo {author} {\bibfnamefont {D.~M.}\ \bibnamefont {Stamper-Kurn}}, \
  and\ \bibinfo {author} {\bibfnamefont {A.}~\bibnamefont {Vishwanath}},\
  }\href {https://doi.org/10.1038/s41567-017-0030-7} {\bibfield  {journal}
  {\bibinfo  {journal} {Nat. Phys.}\ }\textbf {\bibinfo {volume} {14}},\
  \bibinfo {pages} {405} (\bibinfo {year} {2018})}\BibitemShut {NoStop}%
\bibitem [{\citenamefont {F\aa{}k}\ \emph {et~al.}(2012)\citenamefont
  {F\aa{}k}, \citenamefont {Kermarrec}, \citenamefont {Messio}, \citenamefont
  {Bernu}, \citenamefont {Lhuillier}, \citenamefont {Bert}, \citenamefont
  {Mendels}, \citenamefont {Koteswararao}, \citenamefont {Bouquet},
  \citenamefont {Ollivier}, \citenamefont {Hillier}, \citenamefont {Amato},
  \citenamefont {Colman},\ and\ \citenamefont {Wills}}]{Fak2012}%
  \BibitemOpen
  \bibfield  {author} {\bibinfo {author} {\bibfnamefont {B.}~\bibnamefont
  {F\aa{}k}}, \bibinfo {author} {\bibfnamefont {E.}~\bibnamefont {Kermarrec}},
  \bibinfo {author} {\bibfnamefont {L.}~\bibnamefont {Messio}}, \bibinfo
  {author} {\bibfnamefont {B.}~\bibnamefont {Bernu}}, \bibinfo {author}
  {\bibfnamefont {C.}~\bibnamefont {Lhuillier}}, \bibinfo {author}
  {\bibfnamefont {F.}~\bibnamefont {Bert}}, \bibinfo {author} {\bibfnamefont
  {P.}~\bibnamefont {Mendels}}, \bibinfo {author} {\bibfnamefont
  {B.}~\bibnamefont {Koteswararao}}, \bibinfo {author} {\bibfnamefont
  {F.}~\bibnamefont {Bouquet}}, \bibinfo {author} {\bibfnamefont
  {J.}~\bibnamefont {Ollivier}}, \bibinfo {author} {\bibfnamefont {A.~D.}\
  \bibnamefont {Hillier}}, \bibinfo {author} {\bibfnamefont {A.}~\bibnamefont
  {Amato}}, \bibinfo {author} {\bibfnamefont {R.~H.}\ \bibnamefont {Colman}}, \
  and\ \bibinfo {author} {\bibfnamefont {A.~S.}\ \bibnamefont {Wills}},\ }\href
  {\doibase 10.1103/PhysRevLett.109.037208} {\bibfield  {journal} {\bibinfo
  {journal} {Phys. Rev. Lett.}\ }\textbf {\bibinfo {volume} {109}},\ \bibinfo
  {pages} {037208} (\bibinfo {year} {2012})}\BibitemShut {NoStop}%
\bibitem [{\citenamefont {Bieri}\ \emph {et~al.}(2015)\citenamefont {Bieri},
  \citenamefont {Messio}, \citenamefont {Bernu},\ and\ \citenamefont
  {Lhuillier}}]{Bieri2015}%
  \BibitemOpen
  \bibfield  {author} {\bibinfo {author} {\bibfnamefont {S.}~\bibnamefont
  {Bieri}}, \bibinfo {author} {\bibfnamefont {L.}~\bibnamefont {Messio}},
  \bibinfo {author} {\bibfnamefont {B.}~\bibnamefont {Bernu}}, \ and\ \bibinfo
  {author} {\bibfnamefont {C.}~\bibnamefont {Lhuillier}},\ }\href
  {http://link.aps.org/doi/10.1103/PhysRevB.92.060407} {\bibfield  {journal}
  {\bibinfo  {journal} {Phys. Rev. B}\ }\textbf {\bibinfo {volume} {92}},\
  \bibinfo {pages} {060407(R)} (\bibinfo {year} {2015})}\BibitemShut {NoStop}%
\bibitem [{\citenamefont {Pereira}\ and\ \citenamefont
  {Bieri}(2018)}]{Pereira2018}%
  \BibitemOpen
  \bibfield  {author} {\bibinfo {author} {\bibfnamefont {R.~G.}\ \bibnamefont
  {Pereira}}\ and\ \bibinfo {author} {\bibfnamefont {S.}~\bibnamefont
  {Bieri}},\ }\href {\doibase 10.21468/SciPostPhys.4.1.004} {\bibfield
  {journal} {\bibinfo  {journal} {SciPost Phys.}\ }\textbf {\bibinfo {volume}
  {4}},\ \bibinfo {pages} {004} (\bibinfo {year} {2018})}\BibitemShut {NoStop}%
\bibitem [{\citenamefont {Bauer}\ \emph {et~al.}(2019)\citenamefont {Bauer},
  \citenamefont {Keller}, \citenamefont {Trebst},\ and\ \citenamefont
  {Ludwig}}]{Bauer2019}%
  \BibitemOpen
  \bibfield  {author} {\bibinfo {author} {\bibfnamefont {B.}~\bibnamefont
  {Bauer}}, \bibinfo {author} {\bibfnamefont {B.~P.}\ \bibnamefont {Keller}},
  \bibinfo {author} {\bibfnamefont {S.}~\bibnamefont {Trebst}}, \ and\ \bibinfo
  {author} {\bibfnamefont {A.~W.~W.}\ \bibnamefont {Ludwig}},\ }\href
  {https://link.aps.org/doi/10.1103/PhysRevB.99.035155} {\bibfield  {journal}
  {\bibinfo  {journal} {Phys. Rev. B}\ }\textbf {\bibinfo {volume} {99}},\
  \bibinfo {pages} {035155} (\bibinfo {year} {2019})}\BibitemShut {NoStop}%
\bibitem [{\citenamefont {Fradkin}(2013)}]{FradkinBook}%
  \BibitemOpen
  \bibfield  {author} {\bibinfo {author} {\bibfnamefont {E.}~\bibnamefont
  {Fradkin}},\ }\href@noop {} {\emph {\bibinfo {title} {Field Theories of
  Condensed Matter Physics}}},\ Field Theories of Condensed Matter Physics\
  (\bibinfo  {publisher} {Cambridge University Press},\ \bibinfo {year}
  {2013})\BibitemShut {NoStop}%
\bibitem [{\citenamefont {Cardy}(1986)}]{Cardy1986}%
  \BibitemOpen
  \bibfield  {author} {\bibinfo {author} {\bibfnamefont {J.~L.}\ \bibnamefont
  {Cardy}},\ }\href
  {http://www.sciencedirect.com/science/article/pii/0550321386905961}
  {\bibfield  {journal} {\bibinfo  {journal} {Nucl. Phys. B}\ }\textbf
  {\bibinfo {volume} {275}},\ \bibinfo {pages} {200 } (\bibinfo {year}
  {1986})}\BibitemShut {NoStop}%
\bibitem [{\citenamefont {Oshikawa}, \citenamefont {Chamon},\ and\
  \citenamefont {Affleck}(2006)}]{Oshikawa2006}%
  \BibitemOpen
  \bibfield  {author} {\bibinfo {author} {\bibfnamefont {M.}~\bibnamefont
  {Oshikawa}}, \bibinfo {author} {\bibfnamefont {C.}~\bibnamefont {Chamon}}, \
  and\ \bibinfo {author} {\bibfnamefont {I.}~\bibnamefont {Affleck}},\ }\href
  {http://stacks.iop.org/1742-5468/2006/i=02/a=P02008} {\bibfield  {journal}
  {\bibinfo  {journal} {J. Stat. Mech.: Theory and Exp.}\ ,\ \bibinfo {pages}
  {P02008}} (\bibinfo {year} {2006})}\BibitemShut {NoStop}%
\bibitem [{\citenamefont {Kane}, \citenamefont {Mukhopadhyay},\ and\
  \citenamefont {Lubensky}(2002)}]{Kane2002}%
  \BibitemOpen
  \bibfield  {author} {\bibinfo {author} {\bibfnamefont {C.~L.}\ \bibnamefont
  {Kane}}, \bibinfo {author} {\bibfnamefont {R.}~\bibnamefont {Mukhopadhyay}},
  \ and\ \bibinfo {author} {\bibfnamefont {T.~C.}\ \bibnamefont {Lubensky}},\
  }\href {https://link.aps.org/doi/10.1103/PhysRevLett.88.036401} {\bibfield
  {journal} {\bibinfo  {journal} {Phys. Rev. Lett.}\ }\textbf {\bibinfo
  {volume} {88}},\ \bibinfo {pages} {036401} (\bibinfo {year}
  {2002})}\BibitemShut {NoStop}%
\bibitem [{\citenamefont {Fuji}\ and\ \citenamefont
  {Furusaki}(2019)}]{Fuji2019}%
  \BibitemOpen
  \bibfield  {author} {\bibinfo {author} {\bibfnamefont {Y.}~\bibnamefont
  {Fuji}}\ and\ \bibinfo {author} {\bibfnamefont {A.}~\bibnamefont
  {Furusaki}},\ }\href {https://link.aps.org/doi/10.1103/PhysRevB.99.035130}
  {\bibfield  {journal} {\bibinfo  {journal} {Phys. Rev. B}\ }\textbf {\bibinfo
  {volume} {99}},\ \bibinfo {pages} {035130} (\bibinfo {year}
  {2019})}\BibitemShut {NoStop}%
\bibitem [{\citenamefont {Tikhonov}\ and\ \citenamefont
  {Shimshoni}(2019)}]{Tikhonov2019}%
  \BibitemOpen
  \bibfield  {author} {\bibinfo {author} {\bibfnamefont {P.}~\bibnamefont
  {Tikhonov}}\ and\ \bibinfo {author} {\bibfnamefont {E.}~\bibnamefont
  {Shimshoni}},\ }\href {https://link.aps.org/doi/10.1103/PhysRevB.99.174429}
  {\bibfield  {journal} {\bibinfo  {journal} {Phys. Rev. B}\ }\textbf {\bibinfo
  {volume} {99}},\ \bibinfo {pages} {174429} (\bibinfo {year}
  {2019})}\BibitemShut {NoStop}%
\bibitem [{\citenamefont {Song}\ \emph {et~al.}(2017)\citenamefont {Song},
  \citenamefont {Huang}, \citenamefont {Fu},\ and\ \citenamefont
  {Hermele}}]{Song2017}%
  \BibitemOpen
  \bibfield  {author} {\bibinfo {author} {\bibfnamefont {H.}~\bibnamefont
  {Song}}, \bibinfo {author} {\bibfnamefont {S.-J.}\ \bibnamefont {Huang}},
  \bibinfo {author} {\bibfnamefont {L.}~\bibnamefont {Fu}}, \ and\ \bibinfo
  {author} {\bibfnamefont {M.}~\bibnamefont {Hermele}},\ }\href
  {https://link.aps.org/doi/10.1103/PhysRevX.7.011020} {\bibfield  {journal}
  {\bibinfo  {journal} {Phys. Rev. X}\ }\textbf {\bibinfo {volume} {7}},\
  \bibinfo {pages} {011020} (\bibinfo {year} {2017})}\BibitemShut {NoStop}%
\bibitem [{\citenamefont {Huang}\ \emph {et~al.}(2017)\citenamefont {Huang},
  \citenamefont {Song}, \citenamefont {Huang},\ and\ \citenamefont
  {Hermele}}]{Huang2017}%
  \BibitemOpen
  \bibfield  {author} {\bibinfo {author} {\bibfnamefont {S.-J.}\ \bibnamefont
  {Huang}}, \bibinfo {author} {\bibfnamefont {H.}~\bibnamefont {Song}},
  \bibinfo {author} {\bibfnamefont {Y.-P.}\ \bibnamefont {Huang}}, \ and\
  \bibinfo {author} {\bibfnamefont {M.}~\bibnamefont {Hermele}},\ }\href
  {https://link.aps.org/doi/10.1103/PhysRevB.96.205106} {\bibfield  {journal}
  {\bibinfo  {journal} {Phys. Rev. B}\ }\textbf {\bibinfo {volume} {96}},\
  \bibinfo {pages} {205106} (\bibinfo {year} {2017})}\BibitemShut {NoStop}%
\bibitem [{\citenamefont {Buccheri}\ \emph {et~al.}(2018)\citenamefont
  {Buccheri}, \citenamefont {Egger}, \citenamefont {Pereira},\ and\
  \citenamefont {Ramos}}]{Buccheri2018}%
  \BibitemOpen
  \bibfield  {author} {\bibinfo {author} {\bibfnamefont {F.}~\bibnamefont
  {Buccheri}}, \bibinfo {author} {\bibfnamefont {R.}~\bibnamefont {Egger}},
  \bibinfo {author} {\bibfnamefont {R.~G.}\ \bibnamefont {Pereira}}, \ and\
  \bibinfo {author} {\bibfnamefont {F.~B.}\ \bibnamefont {Ramos}},\ }\href
  {https://link.aps.org/doi/10.1103/PhysRevB.97.220402} {\bibfield  {journal}
  {\bibinfo  {journal} {Phys. Rev. B}\ }\textbf {\bibinfo {volume} {97}},\
  \bibinfo {pages} {220402(R)} (\bibinfo {year} {2018})}\BibitemShut {NoStop}%
\bibitem [{\citenamefont {Buccheri}\ \emph {et~al.}(2019)\citenamefont
  {Buccheri}, \citenamefont {Egger}, \citenamefont {Pereira},\ and\
  \citenamefont {Ramos}}]{Buccheri2019}%
  \BibitemOpen
  \bibfield  {author} {\bibinfo {author} {\bibfnamefont {F.}~\bibnamefont
  {Buccheri}}, \bibinfo {author} {\bibfnamefont {R.}~\bibnamefont {Egger}},
  \bibinfo {author} {\bibfnamefont {R.~G.}\ \bibnamefont {Pereira}}, \ and\
  \bibinfo {author} {\bibfnamefont {F.~B.}\ \bibnamefont {Ramos}},\ }\href
  {http://www.sciencedirect.com/science/article/pii/S0550321319300677}
  {\bibfield  {journal} {\bibinfo  {journal} {Nucl. Phys. B}\ }\textbf
  {\bibinfo {volume} {941}},\ \bibinfo {pages} {794 } (\bibinfo {year}
  {2019})}\BibitemShut {NoStop}%
 \bibitem{foot1}
 The vectors $\boldsymbol\delta_\alpha$ are given by
  $\boldsymbol\delta_1=\ell (\sqrt3,0)$, $\boldsymbol\delta_2=\ell (-\frac{\sqrt3}2,\frac{3}2)$, and $\boldsymbol\delta_3=\ell (-\frac{\sqrt3}2,-\frac{3}2)$.  
\bibitem [{\citenamefont {O'Brien}, \citenamefont {Hermanns},\ and\
  \citenamefont {Trebst}(2016)}]{OBrien2016}%
  \BibitemOpen
  \bibfield  {author} {\bibinfo {author} {\bibfnamefont {K.}~\bibnamefont
  {O'Brien}}, \bibinfo {author} {\bibfnamefont {M.}~\bibnamefont {Hermanns}}, \
  and\ \bibinfo {author} {\bibfnamefont {S.}~\bibnamefont {Trebst}},\ }\href
  {https://link.aps.org/doi/10.1103/PhysRevB.93.085101} {\bibfield  {journal}
  {\bibinfo  {journal} {Phys. Rev. B}\ }\textbf {\bibinfo {volume} {93}},\
  \bibinfo {pages} {085101} (\bibinfo {year} {2016})}\BibitemShut {NoStop}%
\bibitem [{\citenamefont {Bieri}, \citenamefont {Lhuillier},\ and\
  \citenamefont {Messio}(2016)}]{Bieri2016}%
  \BibitemOpen
  \bibfield  {author} {\bibinfo {author} {\bibfnamefont {S.}~\bibnamefont
  {Bieri}}, \bibinfo {author} {\bibfnamefont {C.}~\bibnamefont {Lhuillier}}, \
  and\ \bibinfo {author} {\bibfnamefont {L.}~\bibnamefont {Messio}},\ }\href
  {\doibase 10.1103/PhysRevB.93.094437} {\bibfield  {journal} {\bibinfo
  {journal} {Phys. Rev. B}\ }\textbf {\bibinfo {volume} {93}},\ \bibinfo
  {pages} {094437} (\bibinfo {year} {2016})}\BibitemShut {NoStop}%
\bibitem [{\citenamefont {Claassen}\ \emph {et~al.}(2017)\citenamefont
  {Claassen}, \citenamefont {Jiang}, \citenamefont {Moritz},\ and\
  \citenamefont {Devereaux}}]{Claassen2017}%
  \BibitemOpen
  \bibfield  {author} {\bibinfo {author} {\bibfnamefont {M.}~\bibnamefont
  {Claassen}}, \bibinfo {author} {\bibfnamefont {H.-C.}\ \bibnamefont {Jiang}},
  \bibinfo {author} {\bibfnamefont {B.}~\bibnamefont {Moritz}}, \ and\ \bibinfo
  {author} {\bibfnamefont {T.~P.}\ \bibnamefont {Devereaux}},\ }\href
  {https://doi.org/10.1038/s41467-017-00876-y} {\bibfield  {journal} {\bibinfo
  {journal} {Nat. Comm.}\ }\textbf {\bibinfo {volume} {8}},\ \bibinfo {pages}
  {1192} (\bibinfo {year} {2017})}\BibitemShut {NoStop}%
\bibitem [{\citenamefont {Endres}\ \emph {et~al.}(2016)\citenamefont {Endres},
  \citenamefont {Bernien}, \citenamefont {Keesling}, \citenamefont {Levine},
  \citenamefont {Anschuetz}, \citenamefont {Krajenbrink}, \citenamefont
  {Senko}, \citenamefont {Vuletic}, \citenamefont {Greiner},\ and\
  \citenamefont {Lukin}}]{Endres2016}%
  \BibitemOpen
  \bibfield  {author} {\bibinfo {author} {\bibfnamefont {M.}~\bibnamefont
  {Endres}}, \bibinfo {author} {\bibfnamefont {H.}~\bibnamefont {Bernien}},
  \bibinfo {author} {\bibfnamefont {A.}~\bibnamefont {Keesling}}, \bibinfo
  {author} {\bibfnamefont {H.}~\bibnamefont {Levine}}, \bibinfo {author}
  {\bibfnamefont {E.~R.}\ \bibnamefont {Anschuetz}}, \bibinfo {author}
  {\bibfnamefont {A.}~\bibnamefont {Krajenbrink}}, \bibinfo {author}
  {\bibfnamefont {C.}~\bibnamefont {Senko}}, \bibinfo {author} {\bibfnamefont
  {V.}~\bibnamefont {Vuletic}}, \bibinfo {author} {\bibfnamefont
  {M.}~\bibnamefont {Greiner}}, \ and\ \bibinfo {author} {\bibfnamefont
  {M.~D.}\ \bibnamefont {Lukin}},\ }\href
  {http://science.sciencemag.org/content/354/6315/1024} {\bibfield  {journal}
  {\bibinfo  {journal} {Science}\ }\textbf {\bibinfo {volume} {354}},\ \bibinfo
  {pages} {1024} (\bibinfo {year} {2016})}\BibitemShut {NoStop}%
\bibitem [{\citenamefont {{Choi}}\ \emph {et~al.}(2019)\citenamefont {{Choi}},
  \citenamefont {{Lorente}}, \citenamefont {{Wiebe}}, \citenamefont {{von
  Bergmann}}, \citenamefont {{Otte}},\ and\ \citenamefont
  {{Heinrich}}}]{Choi2019}%
  \BibitemOpen
  \bibfield  {author} {\bibinfo {author} {\bibfnamefont {D.-J.}\ \bibnamefont
  {{Choi}}}, \bibinfo {author} {\bibfnamefont {N.}~\bibnamefont {{Lorente}}},
  \bibinfo {author} {\bibfnamefont {J.}~\bibnamefont {{Wiebe}}}, \bibinfo
  {author} {\bibfnamefont {K.}~\bibnamefont {{von Bergmann}}}, \bibinfo
  {author} {\bibfnamefont {A.~F.}\ \bibnamefont {{Otte}}}, \ and\ \bibinfo
  {author} {\bibfnamefont {A.~J.}\ \bibnamefont {{Heinrich}}},\ }\href
  {https://arxiv.org/abs/1904.09941} {\bibfield  {journal} {\bibinfo  {journal}
  {arXiv:1904.09941}\ } (\bibinfo {year} {2019})}\BibitemShut {NoStop}%
\bibitem [{\citenamefont {Wang}\ \emph {et~al.}(2019)\citenamefont {Wang},
  \citenamefont {Song}, \citenamefont {Feng}, \citenamefont {Cai},
  \citenamefont {Xu}, \citenamefont {Deng}, \citenamefont {Li}, \citenamefont
  {Zheng}, \citenamefont {Zhu}, \citenamefont {Wang}, \citenamefont {Zhu},\
  and\ \citenamefont {Scully}}]{Wang2019}%
  \BibitemOpen
  \bibfield  {author} {\bibinfo {author} {\bibfnamefont {D.-W.}\ \bibnamefont
  {Wang}}, \bibinfo {author} {\bibfnamefont {C.}~\bibnamefont {Song}}, \bibinfo
  {author} {\bibfnamefont {W.}~\bibnamefont {Feng}}, \bibinfo {author}
  {\bibfnamefont {H.}~\bibnamefont {Cai}}, \bibinfo {author} {\bibfnamefont
  {D.}~\bibnamefont {Xu}}, \bibinfo {author} {\bibfnamefont {H.}~\bibnamefont
  {Deng}}, \bibinfo {author} {\bibfnamefont {H.}~\bibnamefont {Li}}, \bibinfo
  {author} {\bibfnamefont {D.}~\bibnamefont {Zheng}}, \bibinfo {author}
  {\bibfnamefont {X.}~\bibnamefont {Zhu}}, \bibinfo {author} {\bibfnamefont
  {H.}~\bibnamefont {Wang}}, \bibinfo {author} {\bibfnamefont {S.-Y.}\
  \bibnamefont {Zhu}}, \ and\ \bibinfo {author} {\bibfnamefont {M.~O.}\
  \bibnamefont {Scully}},\ }\href {https://doi.org/10.1038/s41567-018-0400-9}
  {\bibfield  {journal} {\bibinfo  {journal} {Nat. Phys.}\ }\textbf {\bibinfo
  {volume} {15}},\ \bibinfo {pages} {382} (\bibinfo {year} {2019})}\BibitemShut
  {NoStop}%
\bibitem [{\citenamefont {Yang}, \citenamefont {Paramekanti},\ and\
  \citenamefont {Kim}(2010)}]{Yang2010}%
  \BibitemOpen
  \bibfield  {author} {\bibinfo {author} {\bibfnamefont {B.-J.}\ \bibnamefont
  {Yang}}, \bibinfo {author} {\bibfnamefont {A.}~\bibnamefont {Paramekanti}}, \
  and\ \bibinfo {author} {\bibfnamefont {Y.~B.}\ \bibnamefont {Kim}},\ }\href
  {https://link.aps.org/doi/10.1103/PhysRevB.81.134418} {\bibfield  {journal}
  {\bibinfo  {journal} {Phys. Rev. B}\ }\textbf {\bibinfo {volume} {81}},\
  \bibinfo {pages} {134418} (\bibinfo {year} {2010})}\BibitemShut {NoStop}%
\bibitem [{\citenamefont {Jahromi}\ and\ \citenamefont
  {Or\'us}(2018)}]{Jahromi2018}%
  \BibitemOpen
  \bibfield  {author} {\bibinfo {author} {\bibfnamefont {S.~S.}\ \bibnamefont
  {Jahromi}}\ and\ \bibinfo {author} {\bibfnamefont {R.}~\bibnamefont
  {Or\'us}},\ }\href {https://link.aps.org/doi/10.1103/PhysRevB.98.155108}
  {\bibfield  {journal} {\bibinfo  {journal} {Phys. Rev. B}\ }\textbf {\bibinfo
  {volume} {98}},\ \bibinfo {pages} {155108} (\bibinfo {year}
  {2018})}\BibitemShut {NoStop}%
\bibitem [{\citenamefont {Affleck}\ and\ \citenamefont
  {Haldane}(1987)}]{Affleck1987}%
  \BibitemOpen
  \bibfield  {author} {\bibinfo {author} {\bibfnamefont {I.}~\bibnamefont
  {Affleck}}\ and\ \bibinfo {author} {\bibfnamefont {F.~D.~M.}\ \bibnamefont
  {Haldane}},\ }\href {https://link.aps.org/doi/10.1103/PhysRevB.36.5291}
  {\bibfield  {journal} {\bibinfo  {journal} {Phys. Rev. B}\ }\textbf {\bibinfo
  {volume} {36}},\ \bibinfo {pages} {5291} (\bibinfo {year}
  {1987})}\BibitemShut {NoStop}%
\bibitem [{\citenamefont {Gogolin}, \citenamefont {Nersesyan},\ and\
  \citenamefont {Tsvelik}(2004)}]{Gogolin1998}%
  \BibitemOpen
  \bibfield  {author} {\bibinfo {author} {\bibfnamefont {A.}~\bibnamefont
  {Gogolin}}, \bibinfo {author} {\bibfnamefont {A.}~\bibnamefont {Nersesyan}},
  \ and\ \bibinfo {author} {\bibfnamefont {A.}~\bibnamefont {Tsvelik}},\
  }\href@noop {} {\emph {\bibinfo {title} {Bosonization and Strongly Correlated
  Systems}}}\ (\bibinfo  {publisher} {Cambridge University Press},\ \bibinfo
  {year} {2004})\BibitemShut {NoStop}%
\bibitem [{\citenamefont {Rahmani}\ \emph {et~al.}(2010)\citenamefont
  {Rahmani}, \citenamefont {Hou}, \citenamefont {Feiguin}, \citenamefont
  {Chamon},\ and\ \citenamefont {Affleck}}]{Rahmani2010}%
  \BibitemOpen
  \bibfield  {author} {\bibinfo {author} {\bibfnamefont {A.}~\bibnamefont
  {Rahmani}}, \bibinfo {author} {\bibfnamefont {C.-Y.}\ \bibnamefont {Hou}},
  \bibinfo {author} {\bibfnamefont {A.}~\bibnamefont {Feiguin}}, \bibinfo
  {author} {\bibfnamefont {C.}~\bibnamefont {Chamon}}, \ and\ \bibinfo {author}
  {\bibfnamefont {I.}~\bibnamefont {Affleck}},\ }\href
  {https://link.aps.org/doi/10.1103/PhysRevLett.105.226803} {\bibfield
  {journal} {\bibinfo  {journal} {Phys. Rev. Lett.}\ }\textbf {\bibinfo
  {volume} {105}},\ \bibinfo {pages} {226803} (\bibinfo {year}
  {2010})}\BibitemShut {NoStop}%
\bibitem [{\citenamefont {Chalker}\ and\ \citenamefont
  {Coddington}(1988)}]{Chalker1988}%
  \BibitemOpen
  \bibfield  {author} {\bibinfo {author} {\bibfnamefont {J.~T.}\ \bibnamefont
  {Chalker}}\ and\ \bibinfo {author} {\bibfnamefont {P.~D.}\ \bibnamefont
  {Coddington}},\ }\href {https://doi.org/10.1088%2F0022-3719%2F21%2F14%2F008}
  {\bibfield  {journal} {\bibinfo  {journal} {J. Phys. C: Solid State Phys.}\
  }\textbf {\bibinfo {volume} {21}},\ \bibinfo {pages} {2665} (\bibinfo {year}
  {1988})}\BibitemShut {NoStop}%
\bibitem [{\citenamefont {Kitaev}(2003)}]{Kitaev2003}%
  \BibitemOpen
  \bibfield  {author} {\bibinfo {author} {\bibfnamefont {A.}~\bibnamefont
  {Kitaev}},\ }\href {\doibase https://doi.org/10.1016/S0003-4916(02)00018-0}
  {\bibfield  {journal} {\bibinfo  {journal} {Annals of Physics}\ }\textbf
  {\bibinfo {volume} {303}},\ \bibinfo {pages} {2 } (\bibinfo {year}
  {2003})}\BibitemShut {NoStop}%
\bibitem [{\citenamefont {White}(1992)}]{White1992}%
  \BibitemOpen
  \bibfield  {author} {\bibinfo {author} {\bibfnamefont {S.~R.}\ \bibnamefont
  {White}},\ }\href {https://link.aps.org/doi/10.1103/PhysRevLett.69.2863}
  {\bibfield  {journal} {\bibinfo  {journal} {Phys. Rev. Lett.}\ }\textbf
  {\bibinfo {volume} {69}},\ \bibinfo {pages} {2863} (\bibinfo {year}
  {1992})}\BibitemShut {NoStop}%
\bibitem [{\citenamefont {{Chepiga}}\ and\ \citenamefont
  {{White}}(2019)}]{Chepiga2019}%
  \BibitemOpen
  \bibfield  {author} {\bibinfo {author} {\bibfnamefont {N.}~\bibnamefont
  {{Chepiga}}}\ and\ \bibinfo {author} {\bibfnamefont {S.~R.}\ \bibnamefont
  {{White}}},\ }\href {https://arxiv.org/abs/1903.00432} {\bibfield  {journal}
  {\bibinfo  {journal} {arXiv:1903.00432}\ } (\bibinfo {year}
  {2019})}\BibitemShut {NoStop}%
\bibitem [{Note1()}]{Note1}%
  \BibitemOpen
  \bibinfo {note} {See the accompanying Supplementary Material, where we 
  provide details about the bosonization scheme and
  present additional DMRG results.}\BibitemShut {Stop}%
\bibitem [{\citenamefont {Bergholtz}\ and\ \citenamefont
  {Karlhede}(2005)}]{Bergholtz2005}%
  \BibitemOpen
  \bibfield  {author} {\bibinfo {author} {\bibfnamefont {E.~J.}\ \bibnamefont
  {Bergholtz}}\ and\ \bibinfo {author} {\bibfnamefont {A.}~\bibnamefont
  {Karlhede}},\ }\href {\doibase 10.1103/PhysRevLett.94.026802} {\bibfield
  {journal} {\bibinfo  {journal} {Phys. Rev. Lett.}\ }\textbf {\bibinfo
  {volume} {94}},\ \bibinfo {pages} {026802} (\bibinfo {year}
  {2005})}\BibitemShut {NoStop}%
 
\end{thebibliography}

\end{document}